\documentclass[twocolumn,superscriptaddress,aps,preprintnumbers,amsmath,amssymb,sort&compress,nofootinbib
]{revtex4-1}
\usepackage{amsfonts,amsmath,amssymb,ascmac,bm,tensor}
\usepackage{comment}
\usepackage{ifpdf}
\usepackage{slashed}
\usepackage{color}
\usepackage[mathscr]{eucal}
\usepackage[utopia]{mathdesign}
\usepackage[utf8]{inputenc}

\usepackage{cancel}

\ifpdf
  \usepackage{graphicx}     %   usepackage without driver option
  \usepackage[bookmarksopen,colorlinks=true,linkcolor=bblue,citecolor=bblue,urlcolor=ppink]{hyperref}
\else     % For (p)LaTeX + dvipdfmx
  \usepackage[dvipdfmx]{graphicx}     %   usepackage with driver option
  \usepackage[dvipdfmx,bookmarksopen,colorlinks=true,linkcolor=bblue,citecolor=ppink,urlcolor=darkred]{hyperref}
\fi

\definecolor{red}{rgb}{1,0,0}
\definecolor{darkred}{rgb}{0.6,0,0}
\definecolor{darkgreen}{rgb}{0.992447,0.623778,0.034597}
\definecolor{ppink}{rgb}{1,0.4,0.4}
\definecolor{bblue}{rgb}{0.284602,0.317763,0.963947}
\newcommand{\vev}[1]{ \left< {#1} \right> }

\newcommand{\prn}[1]{\left( {#1} \right)}
\newcommand{\com}[1]{\left[ {#1} \right]}
\newcommand{\der}{\partial}
\newcommand{\dd}{\mathrm{d}}
\newcommand{\Mpl}{M_{\rm Pl}}

\newcommand{\abs}[1]{\left\vert {#1} \right\vert}

\def\Mpl{M_{\rm Pl}}

\newcommand{\ee}{\mathrm{e}}

\newcommand{\bae}[1]{\begin{align} #1 \end{align}}
\newcommand{\bce}[1]{\begin{cases} #1 \end{cases}}
\newcommand{\calP}{\mathcal{P}}
\newcommand{\ns}{n_{{}_\text{S}}}
\newcommand{\Si}{\mathrm{Si}}
\newcommand{\Ci}{\mathrm{Ci}}

\makeatletter
\newcommand\footnoteref[1]{\protected@xdef\@thefnmark{\ref{#1}}\@footnotemark}
\makeatother

\begin{document}

%%%%%%%%%%%%%%%%%%%%%%%%%%%
%%%%%%%%%%% Title %%%%%%%%%%%
%%%%%%%%%%%%%%%%%%%%%%%%%%%

%%paper
\title{
Inflationary primordial black holes for the LIGO gravitational wave events\\
and pulsar timing array experiments
}
\author{Keisuke Inomata}
\affiliation{ICRR, University of Tokyo, Kashiwa, 277-8582, Japan}
\affiliation{Kavli IPMU (WPI), UTIAS, University of Tokyo, Kashiwa, 277-8583, Japan}
\author{Masahiro Kawasaki}
\affiliation{ICRR, University of Tokyo, Kashiwa, 277-8582, Japan}
\affiliation{Kavli IPMU (WPI), UTIAS, University of Tokyo, Kashiwa, 277-8583, Japan}
\author{Kyohei Mukaida}
\affiliation{Kavli IPMU (WPI), UTIAS, University of Tokyo, Kashiwa, 277-8583, Japan}
\author{Yuichiro Tada}
\affiliation{ICRR, University of Tokyo, Kashiwa, 277-8582, Japan}
\affiliation{Kavli IPMU (WPI), UTIAS, University of Tokyo, Kashiwa, 277-8583, Japan}
\author{Tsutomu T.~Yanagida}
\affiliation{Kavli IPMU (WPI), UTIAS, University of Tokyo, Kashiwa, 277-8583, Japan}

\begin{abstract}
\noindent
Primordial black holes (PBHs) are one of the candidates to explain the gravitational wave (GW) signals
observed by the LIGO detectors.
Among several phenomena in the early Universe,
cosmic inflation is a major example to generate PBHs from large primordial density perturbations.
In this paper, we discuss the possibility to interpret the observed GW events as mergers of PBHs
which are produced by cosmic inflation.
The primordial curvature perturbation should be large enough to produce a sizable amount of PBHs and thus we have several other probes to test this scenario.
We point out that the current pulsar timing array (PTA) experiments already put severe constraints
on GWs generated via the second-order effects,
and that the observation of the cosmic microwave background (CMB) puts severe restriction on its $\mu$ distortion.
In particular, it is found that
the scalar power spectrum should have a very sharp peak at 
$k \sim 10^{6}$\,Mpc$^{-1}$ to fulfill the required abundance of PBHs
while evading constraints from the PTA experiments together with the $\mu$ distortion.
We propose a mechanism which can realize such a sharp peak.
In the future,
simple inflation models that generate PBHs via almost Gaussian fluctuations %of slowly rolling inflaton
could be probed/excluded.

\end{abstract}

\date{\today}
\maketitle
\preprint{IPMU16-0172}

%%%%%%%%%%%%%%%%%%%%%%%%%%%%%%%%%
%%%%%%%%%%% Introduction %%%%%%%%%%%
%%%%%%%%%%%%%%%%%%%%%%%%%%%%%%%%%
\section{Introduction}\label{sec: intro}

The first detection of a gravitational wave signal was announced
by the LIGO-Virgo Collaboration~\cite{Abbott:2016blz}.
This event, GW150914, comes from a merger of two black holes (BHs) whose masses are
$36^{+5}_{-4}$\,$M_\odot$ and $29^{+4}_{-4}$\,$M_\odot$.
Later, another BH-BH merger event, GW151226, was reported~\cite{Abbott:2016nmj},
whose masses are $14.2^{+8.3}_{-3.7}$\,$M_\odot$ and $7.5^{+2.3}_{-2.3}$\,$M_\odot$,
and also there is a possible candidate of a BH binary, LVT151012.
The event rate of BH-BH merger is estimated as
$9$--$240$\,Gpc$^{-3}$\,yr$^{-1}$ by LIGO-Virgo Collaboration~\cite{TheLIGOScientific:2016pea}.
These remarkable results motivate us to explore the origin of those BHs.

Primordial Black Holes (PBHs) are one of the candidates
to explain observed gravitational wave (GW) events~\cite{Bird:2016dcv,Clesse:2016vqa,Sasaki:2016jop,Eroshenko:2016hmn,Carr:2016drx}.
They are formed in the very early stage of the Universe before any astrophysical objects exist
if the  overdense region collapses overcoming the pressure force~\cite{Hawking:1971ei,Carr:1974nx,Carr:1975qj}.
Cosmic inflation is a major example to generate such large density perturbations in the early Universe.
For instance, if the inflaton potential has a plateau regime,
large superhorizon fluctuations are produced 
while the inflaton goes through that regime.
Later, the formation of PBHs can occur at the horizon reentry 
of the perturbed region~\cite{GarciaBellido:1996qt,Kawasaki:1997ju,Yokoyama:1998pt}.

Since the primordial density perturbation for the large scales, $\gtrsim 1 \text{Mpc}$, must satisfy the COBE normalization,
we need a mechanism to enhance it only at the small scale so as to generate PBHs.
However, such an enhanced small scale perturbation could be probed by the spectral distortion of the cosmic microwave background (CMB).
The $\mu$ distortion is one example to probe such a small scale perturbation~\cite{Chluba:2012we,Kohri:2014lza}, which indicates the effective photon chemical potential due to the inefficient photon number changing interactions.
Thus, inflationary PBHs 
with $4 \times 10^2 M_\odot  \lesssim M \lesssim 4 \times 10^{13} M_\odot$ are constrained~\cite{Kohri:2014lza}.\footnote{
\label{fn:gamma}
	Here we adopt the simple analytic analysis~\cite{Carr:1975qj} to relate
	the frequency of the curvature perturbation with the PBH mass
	(\textit{i.e.,} $\gamma = 3^{-3/2}$).
	See {\bf Uncertainties} in Sec.~\ref{sec:pbhligo}.
}

Moreover, a significant amount of tensor perturbation, namely GWs, 
is simultaneously produced by the large scalar perturbation  via the second-order effects,
which could leave observable signals~\cite{Saito:2008jc,Saito:2009jt,Bugaev:2009zh,Bugaev:2010bb}.
Roughly speaking, the energy-momentum tensor of the first order scalar perturbations 
acts as the source term in the equation of motion for GWs~\cite{Ananda:2006af,Baumann:2007zm}.
For an inflation model that can be an origin of PBHs,
these contributions to GWs tend to be much larger than the first order tensor perturbations
from the Bunch-Davies vacuum fluctuations 
because the required scalar perturbations are so large.
Interestingly, current pulsar timing array (PTA) experiments~\cite{Arzoumanian:2015liz,Lentati:2015qwp,Shannon:2015ect} already put severe constraints,
in particular, in the mass range around $0.1 M_\odot \lesssim M \lesssim 10 M_\odot$.\footnoteref{fn:gamma}
This range is relevant for the LIGO events if we would like to interpret them as  PBH mergers.

In this paper,
we discuss the possibility to interpret the LIGO events as GWs from mergers of PBHs
which originate from superhorizon fluctuations during inflation.
We demonstrate that the PTA experiments and the $\mu$ distortion play important roles to probe this scenario,\footnote{
	See \textit{e.g.}, Refs.~\cite{Schutz:2016khr,Clesse:2016ajp} for other constraints from the PTA experiments.
}
taking a double inflation model~\cite{Kawasaki:1997ju,Kawasaki:1998vx,Frampton:2010sw,Kawasaki:2012kn,Kawasaki:2016ijp,Kawasaki:2016pql} as an example.
We will show that 
the scalar power spectrum should have a sharp peak at $k \sim 10^{6}$\,Mpc$^{-1}$.
Most parameters of the above model are excluded, 
since it is difficult to achieve such a steep spectrum owing to the slow roll condition. 
We have proposed a new mechanism to sharpen the spectrum in the double inflation model to elude those constraints.
Several inflation models which might yield the required sharp spectrum are mentioned.

The organization of this paper is as follows.
In Sec.~\ref{sec:pbh_formation}, we review the formation of PBHs and basic formulas.
In Sec.~\ref{sec:pbhligo},
the possibility to interpret the LIGO events as  PBH mergers is investigated.
It is shown that the curvature perturbations are severely constrained by the PTA experiments and the $\mu$ distortion of CMB.
Parametrizing the curvature perturbation in a simple form,
we have shown that the power of the curvature perturbation has to be large (See Fig.~\ref{fig:scatter} for instance).
In Sec.~\ref{sec:double_inf},
we concretely discuss whether PBHs can be an origin of LIGO events,
taking the double inflation model as a concrete example.
Sec.~\ref{sec:conc} is devoted to conclusions.

%%%%%%%%%%%%%%%%%%%%%%%%%%%%%%%%%%%%
\section{PBH formation}
\label{sec:pbh_formation}
%%%%%%%%%%%%%%%%%%%%%%%%%%%%%%%%%%%%

PBHs are formed when an  overdense region overcomes the pressure force and collapses.
Cosmic inflation can generate such an  overdense region
from large curvature perturbations, $\mathcal P_\zeta$.
The formation of PBHs occurs when the perturbed region reenters the horizon.

In the following, we review the formation of PBHs, assuming that
the large primordial density perturbation on a small scale is somehow generated during inflation
without conflicting the observed CMB spectrum.
See Sec.~\ref{sec:double_inf} for a concrete example of the inflation model.
We follow the conventional analysis 
for the formation of PBHs~\cite{Carr:1975qj,Green:1997sz}.
Note that there exist attempts 
to refine it.\footnote{
See \textit{e.g.,} Refs.~\cite{Choptuik:1992jv,Niemeyer:1997mt,Niemeyer:1998ac,Niemeyer:1999ak,Musco:2004ak,Musco:2008hv,Musco:2012au,Kuhnel:2015vtw} 
and a recent review~\cite{Gundlach:2007gc} for the \textit{critical collapse effect}, and \textit{e.g.,} Refs.~\cite{Shibata:1999zs,Harada:2013epa,Nakama:2013ica} for the discussions on the threshold.
}

\paragraph*{\bf PBH mass-frequency relation.}
In the simple analysis,
the mass of PBH is proportional to the horizon mass at the horizon reentry of 
an  overdense region. 
It is estimated by
\begin{align}
	M(k)  
	&= \left. \gamma \rho \frac{4 \pi H^{-3}}{3} \right|_{k = aH}
	\simeq\frac{\gamma M_\text{eq}}{\sqrt{2}} 
	\prn{ \frac{g_{\ast,\text{eq}}}{g_\ast} }^\frac{1}{6}
	\prn{ \frac{k_\text{eq}}{k} }^2  \nonumber \\[.5em]
	&\simeq M_\odot\left(\frac{\gamma}{0.2}\right)\left(\frac{g_*}{10.75}\right)^{-\frac{1}{6}}\left(\frac{k}{1.9\times10^6\,\mathrm{Mpc}^{-1}}\right)^{-2} \label{eq:pbhmass in k} \\
	&\simeq M_\odot
    \left(
    \frac{\gamma}{0.2}
    \right)
    \left(
    \frac{g_\ast}{10.75}
    \right)^{- \frac{1}{6}}
    \left(
    \frac{f}{2.9 \times 10^{-9} \,\textrm{Hz
   }}
    \right)^{-2},
	\label{eq:pbhmass}
\end{align}
where we have assumed that PBHs are formed in the radiation-dominated era,
and the frequency is defined as $f \equiv k /(2\pi)$.
$g_*$ and $g_{*,\text{eq}}=3.36$ are the effective degrees of freedom for energy density at the PBH formation and the matter-radiation equality, respectively. 
For solar-class PBHs, $g_*$ is 
roughly given by $\simeq10.75$.
It can be seen that such PBHs indeed correspond with $f\sim\mathrm{nHz}$ 
where the PTA experiments have their sensitivities.
$M(k)$ represents the mass of PBH that is produced
when the comoving momentum $k$ reenters the horizon, $k = a H$;
$\gamma$ is the ratio between the  PBH mass and the horizon mass;
$k_\text{eq}$ is the comoving momentum that reenters the horizon at the matter-radiation equality;
and  $M_\text{eq}$ is the horizon mass at the matter-radiation equality.
The simple analytic estimation suggests
$\gamma = 3^{-3/2}\simeq0.2$~\cite{Carr:1975qj},
and we take it as a fiducial value
in the following.

\paragraph*{\bf PBH abundance.}
To estimate the abundance of a PBH with a mass $M(k)$,
we need to know the probability that
the density perturbation with a scale $k$ exceeds the threshold $\delta_c$
for a given power spectrum of curvature perturbations, $\mathcal P_\zeta$.
In this paper, we adopt a simple analytic estimation $\delta_c=1/3$~\cite{Carr:1975qj}.
Assuming that the density perturbation follows the Gaussian distribution,
the probability is obtained from\footnote{
	Throughout this paper, we focus on the case where the curvature perturbation is dominated by the Gaussian distribution. For models with enhanced non-Gaussianity, a sizable amount of PBHs can be produced with a smaller/larger amplitude of the curvature perturbation depending on the sign of non-Gaussianity. As discussed in Ref.~\cite{Nakama:2016gzw} (and we will mention later), the GWs from second-order effects can be reduced by this effect.
}
\begin{align}
	\beta (M) =
	\int_{\delta_c}
	\frac{\dd \delta}{\sqrt{2 \pi \sigma^2 (M)}} \, e^{- \frac{\delta^2}{2 \sigma^2 (M)}}
	\simeq 
	\frac{1}{\sqrt{2 \pi}} \frac{1}{\delta_c / \sigma (M)} \, e^{- \frac{\delta_c^2}{2 \sigma^2 (M)}}.
	\label{eq:beta}
\end{align}
Here $\sigma^2 (M)$ is the variance of the coarse-grained density contrast
associated with the PBH mass $M$, which is given by~\cite{Young:2014ana}
\begin{align}
	\sigma^2 (M (k))
	= \int \dd \ln q W^2 (q k^{-1}) \frac{16}{81} \prn{q k^{-1}}^4 \mathcal P_\zeta (q).
	\label{eq:sigma}
\end{align}
$W$ is the window function smoothing over the corresponding scale $k^{-1}$.
We adopt the Gaussian one in the following analysis, $W (x) = e^{-x^2 / 2}$. 
By using $\beta (M)$,
one can estimate the energy density converted to PBH with a mass $M$
at the horizon reentry as $\gamma \beta (M (k)) \rho |_{k = aH}$.
Since PBHs behave as matter after their production,
the energy fraction of PBHs grows until the matter-radiation equality.
Taking this effect into account,
one can estimate the abundance of PBHs within $M$--$M + \mathrm{d}\ln M$ as follows:
\begin{align}
	\frac{\Omega_\text{PBH} (M)}{\Omega_c}
	&\simeq  \left. \frac{\rho_\text{PBH}}{\rho_m} \right|_\text{eq} \frac{\Omega_{m}}{\Omega_c}
	= \prn{\frac{T_M}{T_\text{eq}} \frac{\Omega_{m}}{\Omega_c}} \gamma \beta (M) \\
	& \simeq
	\!\left(\! \frac{\beta (M)}{1.84 \times 10^{-8}} \!\right)
	\!\left(\! \frac{\gamma}{0.2} \!\right)^\frac{3}{2}\!
	\!\left(\! \frac{10.75}{g_{\ast} (T_M)} \!\right)^\frac{1}{4}\!
	\!\left(\! \frac{0.12}{\Omega_ch^2} \!\right)
	\!\left(\! \frac{M}{M_\odot} \!\right)^{-\frac{1}{2}} \hspace{-7pt}, 
	\label{eq:frac}
\end{align}
where $\Omega_{m}$ ($\Omega_c$) is the current density parameter of matter (DM),
and we used the recent value for the DM density $\Omega_ch^2\simeq0.12$~\cite{Ade:2015xua}.
$T_M$ is the temperature at the formation of a PBH with mass $M$,
and $T_\text{eq}$ is the temperature at the matter-radiation equality.
Note that a typical value of the curvature perturbation to produce a sizable amount of PBHs is $\mathcal P_\zeta \sim 0.01$.
The total abundance of PBHs can be expressed as
\begin{align}
	\Omega_{\text{PBH,tot}} = \int \dd \ln M\, \Omega_\text{PBH} (M).
\end{align}
%%

%%%%%%%%%%%%%%%%%%%%%%%%%%%%%%%%%%%%%%%%%%%%
\section{PBH mergers as LIGO events}
\label{sec:pbhligo}
%%%%%%%%%%%%%%%%%%%%%%%%%%%%%%%%%%%%%%%%%%%%

In this section, we explore the possibilities to interpret the LIGO events as mergers of PBHs
which originate from inflationary fluctuations.
In particular, we point out that the current PTA experiments and the $\mu$ distortion of CMB already provide  severe constraints
at the mass scale relevant to the LIGO events.

\paragraph*{\bf Event rate of PBH mergers.}

\begin{figure}
	\centering
	\includegraphics[width=.40\textwidth]{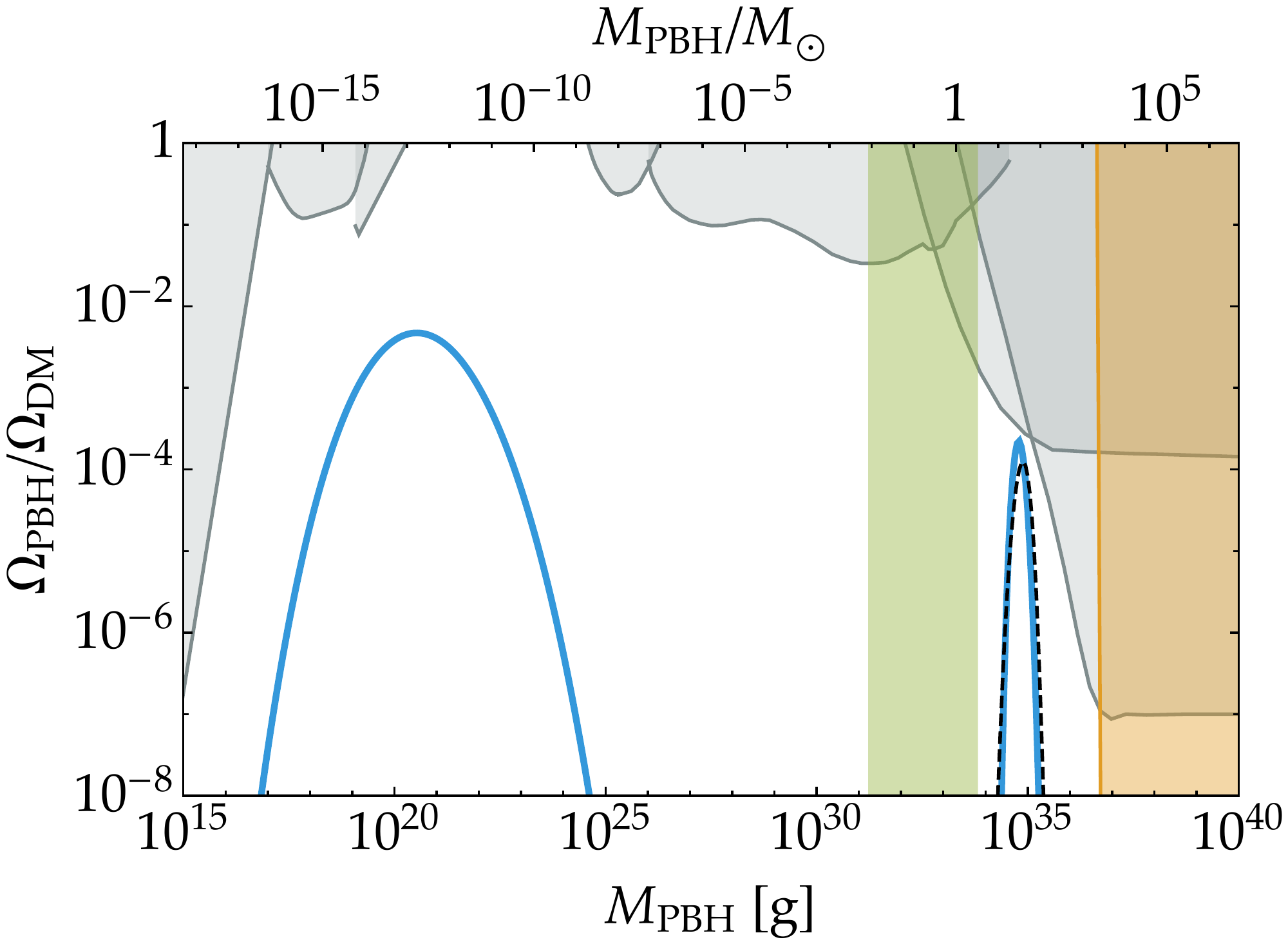}
	\caption{\small
	{\bf Black dashed line}: the PBH mass spectra for parameters given in Eq.~\eqref{eq:not_cancel} ($c_\text{kin}>0$).
	{\bf Cyan solid line}: that for parameters given in Eq.~\eqref{eq:cancel} ($c_\text{kin} < 0$).
	Observationally excluded regions which \textit{do not depend on the production mechanism of PBHs} are represented 
	by {\bf gray-shaded regions}: extragalactic gamma rays from Hawking radiation~\cite{Carr:2009jm},
	femtolensing of known gamma ray bursts~\cite{Barnacka:2012bm}, white dwarfs existing
	in our local galaxy~\cite{Graham:2015apa},
	Kepler micro/millilensing~\cite{Griest:2013esa}, 
	EROS/MACHO microlensing~\cite{Tisserand:2006zx}, and accretion constraints from CMB~\cite{Ricotti:2007au}.
	See also \cite{Carr:2016drx} for a recent summary of observational constraints on PBHs.\footnote{After the submission of this paper, 
	Niikura \textit{et al.}~\cite{Niikura:2017zjd} has shown significant microlensing constraints on $10^{-13}M_\odot\lesssim M\lesssim10^{-6}M_\odot$ with use of the Subaru Hyper Suprime-Cam.}
	We then show the constraints on \textit{inflationary} PBHs in the {\bf orange} and {\bf green} shaded regions:
	the secondary GW constraint with use of the EPTA experiment~\cite{Lentati:2015qwp} and the current $\mu$ distortion constraint $|\mu|<9\times10^{-5}$ by green- and orange-shaded regions 
	for a monochromatic mass spectrum.
	Note that the exact constraints depend on the shape of the power spectrum and therefore the illustrated green/orange constraints are just rough indicators.	
	(See also discussion on uncertainties at the end of Sec.~\ref{sec:pbhligo}.)
	We have provided two sample parameter sets of the double inflation model as an example, which will be discussed in Sec.~\ref{sec:double_inf}.
	Both black dashed [Eq.~\eqref{eq:not_cancel}] and cyan solid [Eq.~\eqref{eq:cancel}] PBH mass spectra seem to avoid these constraints in this figure, but the black dashed one is actually disfavored by PTA constraints as shown in Fig.~\ref{fig:GW}.
	}
	\label{fig:PBH}
\end{figure}

The LIGO-Virgo Collaboration estimated the event rate of BH-mergers as $9$--$240$\,Gpc$^{-3}$\,yr$^{-1}$
(90\,\% C.L.)~\cite{TheLIGOScientific:2016pea}.
As discussed in Ref.~\cite{Sasaki:2016jop}, the required fraction of PBHs is around $\sim 10^{-3}$--$10^{-2}$
so as to reproduce the estimated event rate,
which is disfavored by the CMB observation.
However,  note that there are several uncertainties; for instance,
the above fraction is an order of magnitude estimation under several simplified assumptions
as mentioned in \cite{Sasaki:2016jop},
and the FIRAS constraint could be weaker if one assumes a smaller duty cycle parameter than that in \cite{Ricotti:2007au}
as discussed in \cite{Carr:2016drx}
(though it has been recently claimed that the WMAP3 constraints might be stronger if Planck's results are used instead~\cite{Chen:2016pud}).\footnote{Indeed after the submission of this paper, 
several authors~\cite{Ali-Haimoud:2016mbv,Blum:2016cjs,Horowitz:2016lib} reevaluate the constraints
and find that the conservative ones should be  much weaker than \cite{Ricotti:2007au}.}
Also the observed number of GW events is still quite small.
Therefore, we cannot immediately exclude the possibilities of the PBH-explanation for the LIGO events.
In this sense, it is of quite importance to test this scenario by combining other experiments.
See Fig.~\ref{fig:PBH} for the summary of PBH constraints.

\paragraph*{\bf Induced GWs and PTA experiments.}
Here we summarize the production of GWs via the second-order effects
and show that the PTA experiments could probe GWs associated with the PBHs for the LIGO events.
For the sake of completeness,
we clarify the notation and conventions  in App.~\ref{app:gw_second},
since there is confusion in the literature.

As pointed out in Refs.~\cite{Saito:2008jc,Saito:2009jt,Bugaev:2009zh,Bugaev:2010bb}, a large curvature perturbation can yield a significant amount of GWs that can exceed the usual vacuum contribution.
This is because the curvature perturbations act as a source term in the equation of motion for GWs.
Importantly, those GWs are enhanced at the frequency where the curvature perturbation becomes large.
To understand this intuitively, let us briefly sketch the production of GWs.
GWs are generated when such a large curvature perturbation reenters the horizon. After the production, they are redshifted since they behave as radiation, while the source term decreases much faster.
As a result, the production of GWs is dominated at the horizon reentry of the perturbed region, and the peak of the GWs corresponds to that of the curvature perturbation.

The current density parameter of GWs can be expressed as 
\begin{align}
	\Omega_\text{GW} (\eta_0, k) 
	= \prn{ \frac{a_\text{c}^2 H_\text{c}}{a_0^2 H_0} }^2 \Omega_\text{GW} (\eta_\text{c}, k)
	=  \Omega_{r,0} \Omega_\text{GW} (\eta_c, k),
	\label{eq:app_omega_current}
\end{align}
where $a_0 (=1)$ and $a_c$ are scale factors 
at present and at the time $\Omega_\text{GW}$ becomes constant respectively.
$\eta_c$ represents a conformal time (before the matter-radiation equality $\eta_\text{eq}$) after which the GW density parameter becomes constant.
$\Omega_{r,0}$ is the density parameter of radiation at present.
Thus, all we have to compute is the density parameter at $\eta_c$~(\ref{eq:app_omega_eq}):
\begin{align}
\begin{split}
	\Omega_\text{GW} (\eta_c, k) 
	= \frac{8}{243} \int_0^\infty \dd v \int_{|1 - v|}^{1 + v} \dd u 
	\com{ \frac{4 v^2 - \prn{ 1- u^2 + v^2 }^2 }{4 v u} }^2 &\\
	\quad \times \mathcal P_\zeta (k v) \mathcal P_\zeta (k u) \overline{ I^2 (v, u, k / k_c) },&
\end{split}
\label{eq:app_omega_eq}
\end{align}
Note that the overline indicates the oscillation time average.
Here the integrand kernel is given by~(\ref{eq:app_Ivux})
\begin{align}
	I (v,u,x) &\equiv \int^x_0 \dd \bar x \com{k \frac{a(\bar \eta) \eta_0}{a_0} } \com{k G_{\bm{k}} (\eta , \bar \eta)} 
	f (\tilde{\bm{k}}, \bm{k} - \tilde{\bm{k}}, \bar \eta) \nonumber \\
	\begin{split}
	&= \int^x_0 \dd \bar x \bar x \sin \prn{ x - \bar x } \Big[  
		3 \Psi (v \bar x) \Psi (u \bar x) \\
		& \qquad \quad + \bar x \big\{ \Psi (v \bar x) u \Psi' (u \bar x) + v \Psi' (v \bar x) \Psi (u \bar x)\big\} \\
		& \qquad \qquad \qquad \qquad \qquad \quad + \bar x^2 u v \Psi' (u \bar x) \Psi' (v \bar x)  \Big],
	\end{split}
\end{align} 
where the scalar transfer function is given by~(\ref{eq:app_scalar transfer})
\begin{align}
	\Psi (x) &= \frac{9}{x^2} \com{ \frac{\sin \prn{x/\sqrt{3}}}{x/\sqrt{3}} - \cos \prn{x/\sqrt{3}}}
\end{align}
in the radiation-dominated era.
We have used the following notation, $x \equiv k \eta$, and $\bar x \equiv k \bar \eta$.
This expression is valid for modes that enter the horizon in the radiation-dominated era,
as long as $\eta < \eta_\text{eq}$.

It is instructive to consider 
%the flat 
%\TODO{the delta-function type would be better} power spectrum for the curvature perturbation,
the delta-function spectrum,
\begin{align}
	\mathcal P_\zeta = A \delta ( \log k - \log k_\ast ), 
	\label{eq:}
\end{align}
to get an order of magnitude estimation
of $\Omega_\text{GW}$. One finds
\begin{align}
	\Omega_\text{GW} 
	&= 
	\frac{8}{243} A^2 
	\left[ 1 - \left( \frac{k}{2 k_\ast} \right)^2 \right]^2
	\left( \frac{k_\ast}{k} \right)^2
	\theta \left( 1 - \frac{k}{2 k_\ast} \right)
	\overline{ I^2 \left( \frac{k_\ast}{k}, \frac{k_\ast}{k}, \frac{k}{k_c} \right) }.
	%\Omega_\text{GW}h^2 \sim 10^{-9} \, \prn{ \frac{A}{0.01} }^2.
	\label{eq:omega_gw_order}
\end{align}
Although the function $I$ is complicated,
one can show that this GW spectrum becomes maximal at the momentum of $k_p =  2 k_\ast / \sqrt{3}$.
The resultant GW spectrum may be estimated at the peak momentum as follows (see Eq.~(\ref{eq:app_peakGW})):
\begin{align}\label{eq:peakGW}
	\Omega_\text{GW} (\eta_0, k_p) h^2 
	\sim 1.2 \times 10^{-8} \, \prn{ \frac{\Omega_{r,0}h^2 }{4.2 \times 10^{-5}} }  \prn{\frac{A}{10^{-2}}}^2.
\end{align}
One can see that large curvature perturbation required for PBH formation, 
$\mathcal P_\zeta \sim \mathcal O (0.01)$, yields a substantial amount of GWs,
compared with the current PTA constraints $\Omega_\text{GW}h^2\lesssim10^{-9}$ on $f\sim\mathrm{nHz}$~\cite{Arzoumanian:2015liz,Lentati:2015qwp,Shannon:2015ect} 
as described below in detail.

\begin{figure}
	\centering
	\includegraphics[width=.40\textwidth]{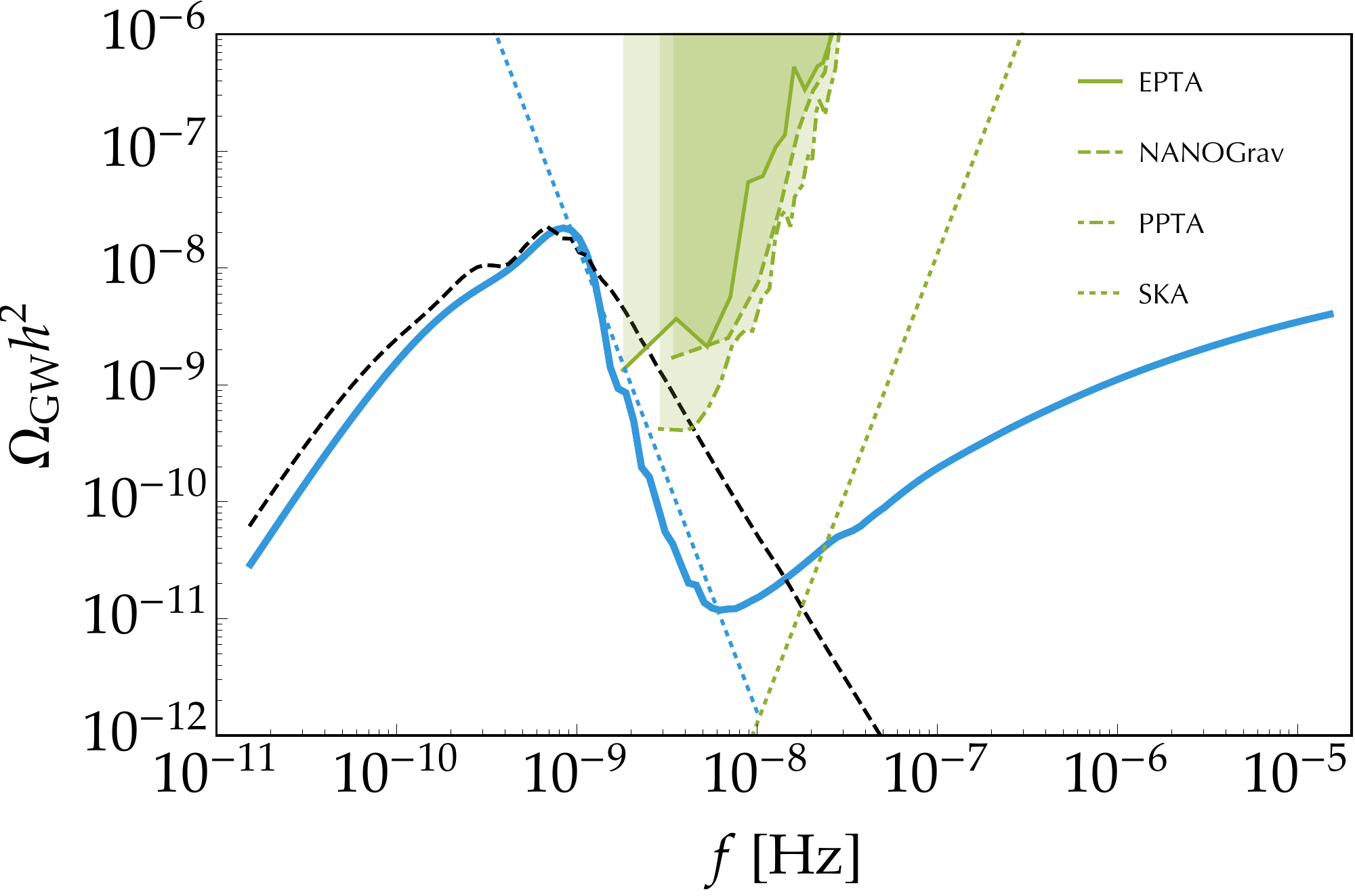}
	\caption{\small
	{\bf Black dashed line}: the induced GW spectra for parameters given in Eq.~\eqref{eq:not_cancel} ($c_\text{kin}>0$).
	{\bf Cyan solid line}: that for parameters given in Eq.~\eqref{eq:cancel} ($c_\text{kin} < 0$).
	We also plot $\Omega_\text{GW} (k) = \Omega_\text{GW, peak} (k_i/k)^4$ in a black dotted line
	for comparison. 
	 The current three severe constraints, \textit{i.e.,}~NANOGrav~\cite{Arzoumanian:2015liz}, EPTA~\cite{Lentati:2015qwp}, 
	 and PPTA~\cite{Shannon:2015ect} are shown in green-shaded regions.
	 The prospect of the SKA sensitivity is shown 
	 in a green dotted line~\cite{Moore:2014lga,Janssen:2014dka}.
	 Although the case of $c_\text{kin} > 0$  is marginal,
	we cannot immediately exclude it because of the uncertainty of the factor $\gamma$.
	A slightly small $\gamma$ is enough to elude this constraint.
	}
	\label{fig:GW}
\end{figure}

The PTA experiments are sensitive at the frequency of $10^{-9}$--$10^{-8}$\,Hz.
As one can see from Eq.~\eqref{eq:pbhmass}, the corresponding PBH mass is around
$\sim 42 \gamma M_\odot$--$0.42 \gamma M_\odot$.
The range depends on the uncertain factor $\gamma$,
which denotes how efficiently the radiation inside the horizon collapses to become a PBH at the horizon crossing.
In the simple analysis, it is estimated as $\gamma = 3^{-3/2} \simeq 0.2$.
The LIGO events ranging from $7.5 M_\odot$ to $36 M_\odot$ could be probed by the PTA experiments
if they come from mergers of PBHs which originate from inflationary fluctuations.

The current PTA experiments already put  restrictions on GWs of the relevant frequency.
Fig.~\ref{fig:GW} shows the present severe three constraints, \textit{i.e.,}~NANOGrav~\cite{Arzoumanian:2015liz}, EPTA~\cite{Lentati:2015qwp}, and PPTA~\cite{Shannon:2015ect}, 
and the future prospect of SKA~\cite{Moore:2014lga,Janssen:2014dka}.
The current constraint reaches $\Omega_\text{GW} h^2 \sim 10^{-9}$
at the frequency of $f \sim (2$--$3) \times 10^{-9}$\,Hz.
It is so severe that the required amount of PBHs, $\Omega_\text{PBH}/\Omega_c \gtrsim \mathcal O(10^{-4})$,
cannot be produced at this frequency
as one can infer from Eqs.~\eqref{eq:beta}, \eqref{eq:sigma}, \eqref{eq:frac}, and \eqref{eq:omega_gw_order}.
Thus, a larger $\gamma$, $\mathcal O (0.1) \ll \gamma \lesssim 1$,  is already ruled out.
We take $\gamma = 3^{-3/2}$ as a fiducial value.

\paragraph*{\bf $\bm \mu$ distortion.}

\begin{figure}
	\centering
	\includegraphics[width=.40\textwidth]{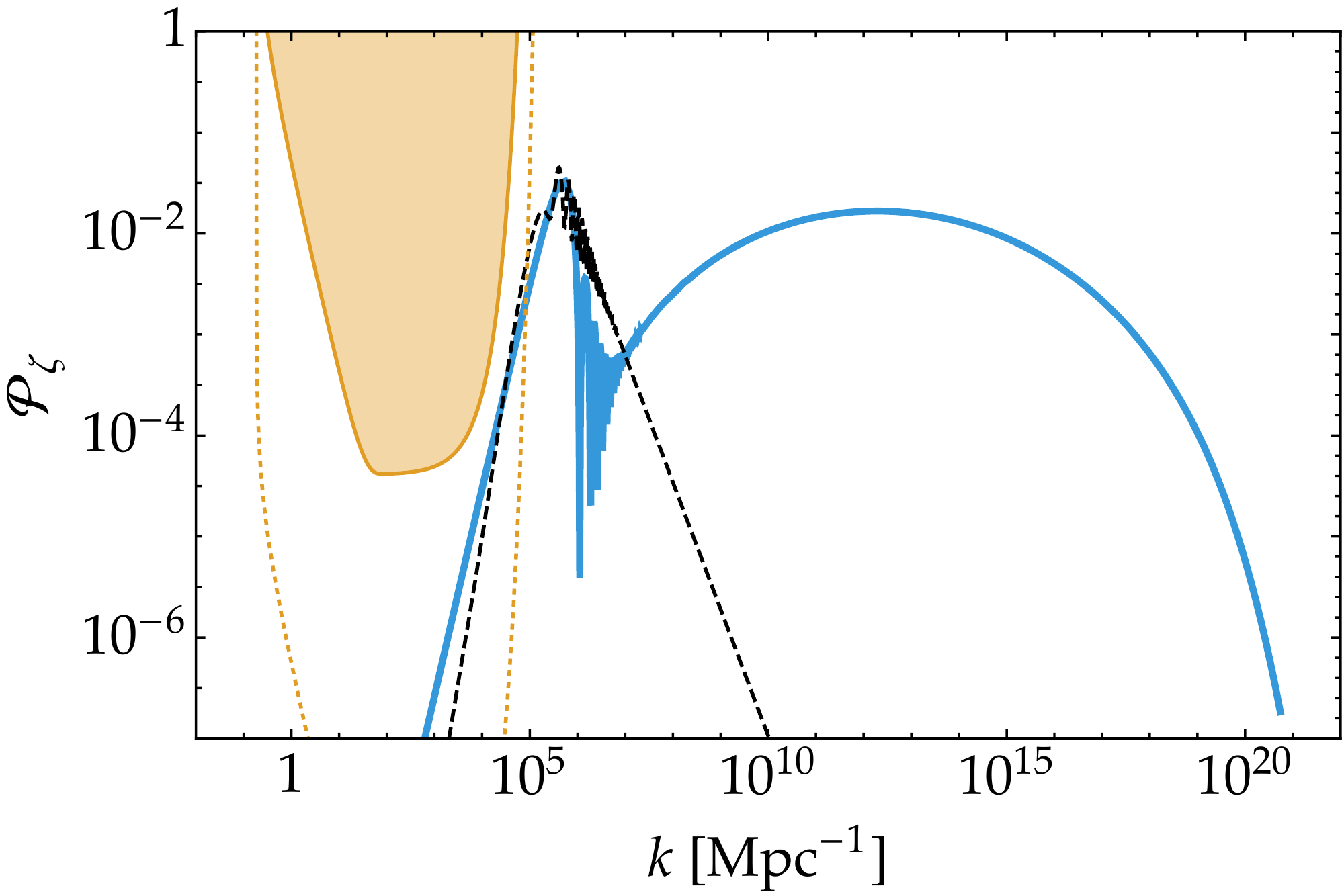}
	\caption{\small
	{\bf Black dashed line}: the scalar power spectra for parameters given in Eq.~\eqref{eq:not_cancel} ($c_\text{kin}>0$).
	{\bf Cyan solid line}: that for parameters given in Eq.~\eqref{eq:cancel} ($c_\text{kin} < 0$).
	Here we show the constraints from the $\mu$ distortion with the current constraint $|\mu|<9\times10^{-5}$~\cite{Fixsen:1996nj}
	by an orange-shaded region and the future prospect $|\mu|<10^{-9}$~\cite{Kogut:2011xw,Andre:2013afa} by a orange dotted line.
	}
	\label{fig:scalar}
\end{figure}

The CMB spectral distortion would be another severe constraint on the primordial small scale perturbations.\footnote{
	See Refs.~\cite{Jeong:2014gna,Nakama:2014vla,Inomata:2016uip} for other probes of small scale perturbations that might be relevant.
	Though they are sensitive to a slightly smaller scale than the $\mu$ distortion which we will discuss,
	their constraining power is similar.
}
In the standard cosmology, the small scale perturbations are erased due to the friction between the photon fluid and the baryon plasma called the Silk damping,
and yield entropy production. While the plasma is soon rethermalized after the entropy production in the sufficiently early Universe,
the rethermalization cannot be completed in the late Universe $z\lesssim 2\times10^6$ due to the inefficiency of the interaction.
Particularly, during $5\times10^4\lesssim z\lesssim 2\times10^6$ corresponding with $50\,\mathrm{Mpc}^{-1}\lesssim k\lesssim10^4\,\mathrm{Mpc}^{-1}$,
the photon-number-changing process such as the double Compton scattering or the electron-positron annihilation becomes inefficient 
and the photon fluid starts to deviate from the Planck distribution though it can reach the kinetic equilibrium. 
Therefore the photon distribution can be written by the general Bose-Einstein distribution whose deviation from the Planck one can be parametrized by the chemical potential $\tilde{\mu}$.
Conventionally the dimensionless $\mu$ parameter $\mu=-\tilde{\mu}/T$ is often used and it is constrained by COBE/FIRAS as~\cite{Fixsen:1996nj}
\bae{
	|\mu|<9\times10^{-5}.
}
On the other hand, the estimated $\mu$ distortion originating from the single $k$ mode perturbation:
\bae{
	\calP_\zeta(k)=A\delta(\log k-\log k_*),
}
can be approximated by~\cite{Chluba:2012we,Kohri:2014lza}
\bae{
	\mu\sim2.2A\left[\exp\left[-\frac{k_*}{5400\,\mathrm{Mpc}^{-1}}\right]-\exp\left[-\left(\frac{k_*}{31.6\,\mathrm{Mpc}^{-1}}\right)^2\right]\right].
}
Combining them, one can obtain the constraint on the primordial perturbations as
\bae{
	A\lesssim10^{-4}, \quad 50\,\mathrm{Mpc}^{-1}\lesssim k\lesssim10^4\,\mathrm{Mpc}^{-1},
}
which is quite severe from the viewpoint of PBH production. We show the explicit form of the constraints in Fig.~\ref{fig:scalar} as an orange region.
Also future space missions such as PIXIE~\cite{Kogut:2011xw} and PRISM~\cite{Andre:2013afa} will improve the constraints on $\mu$ to $\mathcal{O}(10^{-8}\text{--}10^{-9})$.

We also summarize the PBH constraints for monochromatic spectra corresponding with these secondary GW and $\mu$ distortion in Fig.~\ref{fig:PBH} as green- and orange-shaded regions
(for secondary GWs we only consider the peak scale~\eqref{eq:peakGW}.
Note that the exact GW spectrum depends on the shape of the power spectrum due to the momentum convolution, and therefore one has to calculate the concrete GW spectrum for each predicted power spectrum
to check whether the PTA constraints are indeed satisfied or not.

\paragraph*{\bf Constraints summary.}

\begin{figure}
	\centering
	\includegraphics[width=.40\textwidth]{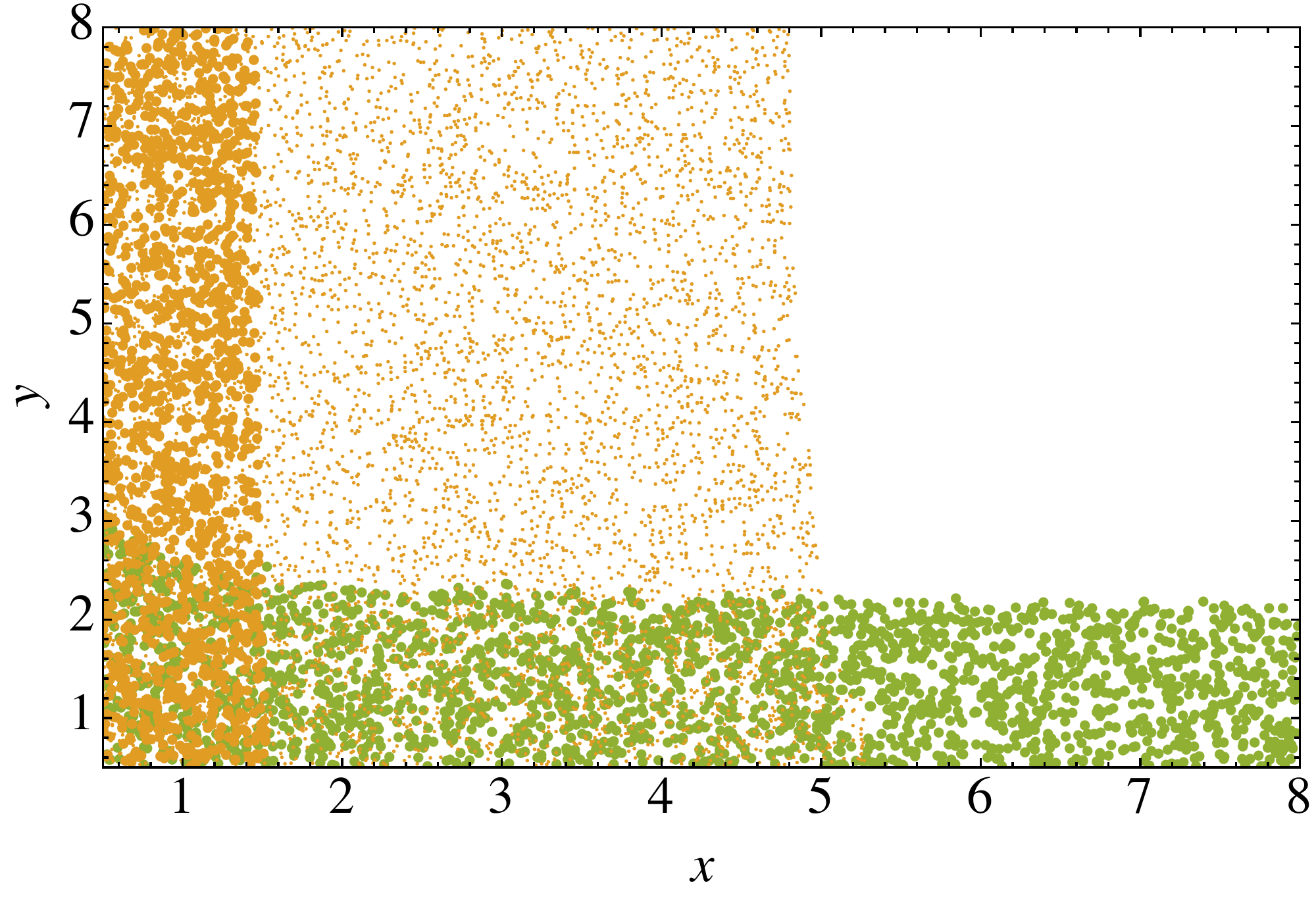}
	\caption{\small
	The excluded parameter regions by the current PTA constraints (large green dots), the current $\mu$ constraint $|\mu|<9\times10^{-5}$ (large orange dots),
	and the future prospect $|\mu|<10^{-9}$ (small orange dots) with the assumption that the PBH mass spectrum becomes maximal as $\Omega_\text{PBH}/\Omega_c=10^{-4}$
	at $30M_\odot$. 
	The future SKA experiment would exclude all regions though we do not plot them to avoid a busy figure.
	}
	\label{fig:scatter}
\end{figure}

To illustrate the impact of the above constraints on the curvature perturbation, we adopt the following parametrization:
\begin{align}
	\mathcal P_\zeta (k)  =
	\begin{cases}
		A \prn{ \cfrac{k}{k_\ast} }^x &\text{for}\,\,\, k < k_\ast,
		\\[1em]
		A \prn{ \cfrac{k}{k_\ast} }^{-y} &\text{for} \,\,\, k_\ast < k,
	\end{cases}
	\label{eq:pzeta_prm}
\end{align}
where $A$ and $k_*$ are determined so that the resultant PBH mass spectrum has a peak as $\Omega_\text{PBH}/\Omega_c=10^{-4}$ at $30M_\odot$,
which slightly depend on $x$ and $y$. $A$ and $k_*$ depend also on the uncertain factor $\gamma$, and for our fiducial value $\gamma=3^{-3/2}$, 
it can be found $k_*\sim3.4\times10^{5}\,\mathrm{Mpc}^{-1}$ from Eq.~(\ref{eq:pbhmass in k}).

Fig.~\ref{fig:scatter} shows the $x$-$y$ regions excluded by the current PTA and $\mu$ constraints and the future $\mu$ constraints.
Basically, the PTA experiments make the spectrum steeper above the peak frequency $k_\ast$, while the $\mu$ distortion does below $k_\ast$.
The allowed region slightly shrinks for a larger abundance of PBH.
The curvature perturbation should be sharp enough to satisfy
$x \gtrsim 1.5$ and $y \gtrsim 2$, which indicates that simple single-field slow-roll inflation models 
are already disfavored as a candidate of LIGO events because the tilt of the power spectrum $x$ or $y$ is slow-roll suppressed in this case.

In the future, the $\mu$ distortion as small as $|\mu|\lesssim10^{-9}$ can be probed, for instance, by PIXIE~\cite{Kogut:2011xw} or PRISM~\cite{Andre:2013afa} and they almost exclude $x\lesssim5$.
Also the SKA experiment may reach $\Omega_\text{GW} h^2 \sim 10^{-15}$ and we checked that the induced GW can be detected with this sensitivity almost irrespective of the tilt of the power spectrum.
Thus, many inflationary models that are responsible for PBHs as the LIGO events could be tested in the future.

\paragraph*{\bf Uncertainties.}
Here possible uncertainties/loopholes in the above estimation are mentioned.

First, let us stress the $\gamma$ dependence again.
As can be seen in Eq.~\eqref{eq:pbhmass}, the correspondence between the GW frequency and the PBH mass has a $\gamma$ uncertainty,
while the simple analytic analysis gives $\gamma = 3^{-3/2}$~\cite{Carr:1975qj}.
Noting that the PTA experiments and $\mu$ distortion constrain the scalar perturbations on $k\sim\sqrt{3}k_p/2\gtrsim10^6\,\mathrm{Mpc}^{-1}$ and $k\lesssim10^5\,\mathrm{Mpc}^{-1}$ respectively,
the allowed range of $\gamma$ may be around $0.02 \lesssim \gamma \lesssim 2$.
Interestingly, the representative value, $\gamma = 3^{-3/2}$, is still allowed.

Second, the required value of the curvature perturbation strongly depends on the threshold value of the density contrast, $\delta_c$ [See Eq.~\eqref{eq:beta}]. 
Some numerical studies indicate a slightly larger value, \textit{e.g.}, $\delta_c\simeq0.45$~\cite{Musco:2004ak,Musco:2008hv,Musco:2012au},
than that used in this paper, which results in slightly more sever constraints.
However note that the precise value of $\gamma$ and $\delta$ is still under discussion. 
Further investigations to narrow down its precise value are quite important.

Finally, inflation models with an enhanced non-Gaussianity can produce a larger/smaller amount of PBHs depending on the sign of non-Gaussianity.
The required amplitude of the curvature perturbation to account for the LIGO events can be smaller/larger.
Hence the constraints on GWs via the second-order effects
and also those on the small scale curvature perturbation
via the $\mu$ distortion can be weaker/stronger~\cite{Nakama:2016gzw}.

\section{Concrete example for the LIGO events avoiding PTA and $\mu$ constraints}
\label{sec:double_inf}

\paragraph*{\bf Double inflation.}
Now we are in a position to discuss a concrete example where sufficient PBHs can be produced for the LIGO events
while current PTA and $\mu$ constraints are avoided.
Here we consider the double inflation model proposed in Ref.~\cite{Kawasaki:1997ju}. This model consists of 
large field inflation as a pre-inflation and a new inflation as a second inflation.
It is attractive, for the pre-inflation dynamically solves
the initial condition problem of the new inflation~\cite{Izawa:1997df}.
Furthermore, the new inflation can yield sizable scalar perturbations at the small scale
without conflicting the observed CMB spectrum,
while the pre-inflation accounts for the scalar perturbations at the large scale observed by Planck~\cite{Ade:2015xua}.

To make our discussion concrete,
we consider the following form for the inflaton potential:
\begin{align}
	V (\phi ,\varphi) 
	=& V_\text{pre} (\phi) 
	+ V_\text{stb} (\phi ,\varphi)
	+ V_\text{new} (\varphi), \\
	V_\text{new} (\varphi)
	=&
	\prn{ v^2 - g \frac{\varphi^n}{\Mpl^{n-2}} }^2 
	- \kappa v^4 \frac{\varphi^2}{2 \Mpl^2} - \varepsilon v^4 \frac{\varphi}{\Mpl}, \\
	V_\text{stb} (\phi,\varphi) =&
	c_\text{pot} \frac{V_\text{pre} (\phi) }{2 \Mpl^2} \varphi^2, 
\end{align}
and as pre-inflation we adopt a quadratic potential $V_\text{pre}=\frac{1}{2}m^2\phi^2$ for simplicity.\footnote{
	Note here that this simple choice is not consistent with Planck's results.
	This is because 
	the $e$-folding number of the pre-inflation for the observable Universe gets smaller than $60$,
	and for the following new inflation continues $\sim 30$ $e$-folds here.
	As a result, the predicted $n_s$ and $r$ are shifted. 
	However almost only the oscillatory behavior of $\phi$ is 
	relevant for small scale perturbations and it can be approximated by a quadratic potential force at the leading order.
	Therefore we assume that our discussions for PBH formation are valid also for other large field inflation 
	than a quadratic potential.
}

Generally speaking, we also expect Planck-suppressed corrections to kinetic terms:
\begin{align}
	\mathcal L_\text{kin} = - \frac{1}{2} \der_\mu \varphi \der^\mu \varphi 
	- \frac{1}{2} \prn{1 - \frac{c_\text{kin}}{2 \Mpl^2} \varphi^2}  \der_\mu \phi \der^\mu \phi  + \cdots.
	\label{eq:kin_term}
\end{align}
Here we only keep terms that are relevant in the following discussion.
See App.~\ref{sec:sugra} for its possible origin.
Here $\phi$ ($\varphi$) represents the inflaton responsible for the chaotic (new) inflation;
$V_\text{pre}$ ($V_\text{new}$) is the potential for the chaotic (new) inflation;
$V_\text{stb}$ stabilizes $\varphi$ during chaotic inflation.
$c_\text{pot}$, $c_\text{kin}$, $g$, $\kappa$, and $\varepsilon$ are dimensionless parameters;
and $v$ is the scale of the new inflation.
Hereafter we assume $c_\text{pot} \sim |c_\text{kin}| \sim \mathcal O (1),\,c_\text{pot}>0,$ and $c_\text{kin}\gtrsim-c_\text{pot}$.

The dynamics of the double inflation proceeds as follows.
First, chaotic inflation takes place while $\phi$ slowly rolls down the potential $V_\text{pre}$.
During this regime,
$\varphi$ is stabilized at $\varphi \sim \varepsilon v^4 \Mpl / (c_\text{pot}V_\text{pre})$.
After the end of chaotic inflation,
the $\phi$ oscillation induces another Hubble induced mass term from Eq.~\eqref{eq:kin_term}, and
$\varphi$ is stabilized at $\varphi \sim \varepsilon v^4 \Mpl / (c \rho_\text{pre})$
with $c = \frac{1}{2}(c_\text{kin} + c_\text{pot})$ until the energy of $\phi$ becomes $\rho_\text{pre} \sim v^4$,
where the new inflation starts.
Hence, the initial field value of $\varphi$ can be estimated as $\varphi_i \sim \varepsilon \Mpl$.
The new inflation occurs until the slow-roll condition fails at
$\varphi_e \sim (v^2 \Mpl^{n-4} / (2n(n-1)g)  )^{1/(n-2)}$.
Finally, $\varphi$ oscillates around the potential minimum located at $\varphi_m \simeq \Mpl (v^2/g \Mpl^2)^{1/n}$
with a mass scale of $m_\varphi \simeq \sqrt{2} n (v^2/\Mpl ) (v^2/g\Mpl^2)^{-1/n}$.
Suppose that $\varphi$ decays via a dimension-five Planck-suppressed operator.\footnote{
	Since the new inflation requires an extremely flat potential,
	it may be natural to assume that it only interacts with the Standard Model sector
	via Planck-suppressed operators.
}
The reheating temperature can be evaluated as
\begin{align}
	T_\text{R} \simeq  0.1\sqrt{\frac{m_\varphi^{3}}{\Mpl}}
	 \simeq
	 0.2n^\frac{3}{2}g^\frac{3}{2n}\Mpl \prn{\frac{v}\Mpl}^{3-\frac{3}{n}}.
	\label{eq:TR}
\end{align}
Throughout this paper,
we take this value of the reheating temperature as a benchmark.

\paragraph*{\bf How to obtain sharp power spectrum.}
There are two ways to realize a sharp peak on the power spectrum of the curvature perturbation in this model (See also cyan solid and black dashed lines in Fig.~\ref{fig:scalar}).
In the following, we discuss both cases in turn.

\textit{In the case of $c_\text{kin} > 0$}, $\varphi$'s perturbation is not affected by $\phi$'s oscillation and the curvature perturbation
generated during new inflation is determined almost only by $\varphi$'s potential form (see App.~\ref{sec:multihorizon}).
Therefore, in this case, one has to take a sizable $\kappa$ and make the power spectrum strongly red-tilted
(note that the tilt of the power spectrum $\ns-1$ is given by $-6\epsilon+2\eta\simeq2\eta$ where $\epsilon=(\Mpl^2/2)(V^\prime/V)^2$ and $\eta=\Mpl^2V^{\prime\prime}/V\simeq-\kappa$ are
the slow-roll parameters).
The power spectrum of curvature perturbations is
\begin{align}
	\mathcal P_\zeta \sim 
	\begin{cases}
	\displaystyle
	\text{max}\left(A\left(\frac{k}{k_i}\right)^{3-2\text{Re}\nu},\,\,
	\left.\mathcal P_\zeta \right|_\text{chaotic}
	\sim 10^{-9} \right)
	&\text{for}\,\,\,k_i \gtrsim k,\\[1em]
	\displaystyle
	A \prn{ \frac{k_i}{k} }^{ 2\kappa }
	&\text{for}\,\,\,k \gtrsim k_i,
	\end{cases}
	\label{eq:power_kappa+app}
\end{align}
where $\nu=\sqrt{9/4-3c_\text{pot}}$ (see App.~\ref{sec:multihorizon}), the amplitude is $A = v^4 / (12 \pi^2 \Mpl^4 \varepsilon^2)$,
and $k_i$ is a comoving momentum that exits the horizon at the onset of the new inflation,
$k_i = a_i H_i$. 
In terms of the parametrization in Eq.~(\ref{eq:pzeta_prm}), $x=3-2\text{Re}\nu\le3$ and $y=2\kappa$.
One can see that a sizable amount of scalar perturbations are generated for $\varepsilon=\alpha v^2/\Mpl^2$ with $\alpha\sim\mathcal{O}(1)$.\footnote{
	There is a lower bound $\alpha > 1/(2 \sqrt{3} \pi)$ to avoid the eternal inflation.
}
Also, the power spectrum has a peak at $k \sim k_i$.
If $k_i$ corresponds to a PBH mass of $\sim 10 M_\odot$,
PBHs responsible for the LIGO events can be produced.
See also Eq.~\eqref{eq:pbhmass}.

The current PTA experiments push the scenario with $c_\text{kin} > 0$ on edge.
This is because the slow-roll condition forces the power not to be so steep $|\kappa | \ll 1$, while the PTA constraints require $y=2\kappa\gtrsim2$ as Fig.~\ref{fig:scatter} shows.
%As a result, it is difficult to achieve a sharp power spectrum.
To demonstrate this, we plot the concrete GW power spectrum for
\begin{align}\label{eq:not_cancel}
	n&=3, \quad \frac{v}{\Mpl}=5 \times 10^{-5}, \quad 
	\kappa=0.76, \quad 
	\alpha=\frac{\varepsilon\Mpl^2}{v^2}=0.74, \nonumber \\ 
	g&=1.13\times10^{-10}, \quad 
	c_\text{pot}=1, \quad 
	c_\text{kin}=0.1,
\end{align}
in Fig.~\ref{fig:GW}, together with the steepest GW spectrum, 
\textit{i.e.,} $\Omega_\text{GW} (k) = \Omega_\text{GW, peak} (k_i/k)^4$.
We have also plotted the scalar power spectrum in Fig.~\ref{fig:scalar}  and the PBH mass spectrum Fig.~\ref{fig:PBH}.
Note that we did not use the slow-roll approximations but numerically solved the full equation of motion 
for the background fields and the linear perturbations to obtain this power spectrum 
(see \textit{e.g.,} Ref.~\cite{Weinberg:2008zzc}).
Although this scenario is marginal due to the current PTA constraints,
we cannot immediately exclude it because of the uncertainty in $\gamma$.

%%%%%%%%%%%%%%%%%%
\begin{comment}
A slightly small $\gamma$ is enough to elude the current constraints.
In the future, the SKA experiment could reach a much smaller value of the GW-density 
close to $\Omega_\text{GW}h^2 \sim 10^{-15}$.
Therefore this scenario will be tested, for the spectrum cannot be so steep to avoid this constraint because
$\Omega_\text{GW} \propto k^{-4 \kappa}$ with $\kappa \ll 1$.
\end{comment}
%%%%%%%%%%%%%%%%%%

In addition, 
for a large $\kappa$, there exists an upper bound on the new inflation scale $v$ of our model.
This is because the $e$-folding number of the new inflation becomes smaller for a larger $\kappa$,
\textit{i.e.,}
\begin{align}
	N_\text{new,tot}
	&\simeq \int_{\varphi_e}^{\varphi_i} \frac{\dd \varphi}{\Mpl^2} \frac{V_\text{new}}{V_\text{new}'}
	\sim \frac{1}{\kappa} \ln \frac{\varphi_e}{\varphi_i}  \nonumber \\
	&\lesssim 
	\prn{\frac{0.8}{\kappa}}
	\prn{31 - \frac{2}{0.8} \ln \frac{v}{10^{13}\,\mathrm{GeV}} 
	- \frac{1}{0.8} \ln \frac{\alpha}{0.8} },
\end{align}
where we used the fact that $\varphi_e$ cannot be much larger than $\Mpl$ for large $\kappa\sim\mathcal{O}(0.1)$ even if the coupling $g$ is very small 
because inflation is ended by the negative mass term $-\kappa v^4\frac{\varphi^2}{2\Mpl^2}$
around $\varphi\sim\Mpl$ in this case. On the other hand,
we need a sufficiently long period of the new inflation to have a heavy PBH
with a mass comparable to $\sim 10 M_\odot$, \textit{i.e.,}
\begin{align}
	N_\text{new} (f) \simeq& 29 - \ln \frac{f}{7 \times 10^{-10}\,\mathrm{Hz}}  \nonumber\\
	&+ \frac{2}{3} \ln \frac{v}{10^{13}\,\mathrm{GeV}}
	+ \frac{1}{3} \ln \frac{T_\text{R}}{10^{5}\, \mathrm{GeV}}.
\end{align}
Combining these two equations and inserting $ \kappa\sim 0.8$, $\alpha \sim 0.8$, and $T_\text{R} \sim 10^5$\,GeV,
one finds a rough upper bound on the new inflation scale: $v \lesssim \mathcal O(10^{13} )$\,GeV.
As a result, the scenario proposed in Ref.~\cite{Kawasaki:2016ijp}
seems to be disfavored in the case of $c_\text{kin} > 0$,
since it requires a high new inflation scale $v \gtrsim 10^{15}\,\text{GeV}$ to avoid
the decay of the metastable electroweak vacuum during the preheating stage after chaotic inflation~\cite{Ema:2016kpf}.\footnote{If one assumes an additional inflationary phase after
the new inflation, this scenario would be realized.}

\textit{In the case of $c_\text{kin}<0$}, the Hubble induced mass for $\varphi$ can be canceled out
in the oscillation phase of $\phi$.
As a result, the modes which are on the superhorizon scale in this phase 
can be steeply enhanced as described in App.~\ref{sec:multihorizon} in detail.
Then it can realize a sharper fall-off than the slow-roll case.
Concretely speaking, the parameter $y$ in Eq.~(\ref{eq:pzeta_prm}) is replaced from $2\kappa$ to $2\kappa+4\text{Re}\nu_\text{osc}$
where $\nu_\text{osc}=\sqrt{9/16-3c_\text{osc}}$ with $c_\text{osc}=(c_\text{pot}+c_\text{kin})/2$.
Therefore if the negative $c_\text{kin}$ sufficiently cancel out $c_\text{pot}$, the power spectrum can steeply fall off as $\propto k^{-3}$
even if $\kappa$ is negligibly small.

Since now $\kappa$ need not be large,
the scenario in Ref.~\cite{Kawasaki:2016ijp} is still viable because the new inflation scale can be high.
Furthermore, if one takes a negative $\kappa$, 
the $\varphi$'s potential can have a flat inflection point,
making a second peak for the scalar power spectrum~\cite{Kawasaki:2016pql}. 
We show the specific example of the parameter set 
\begin{align}\label{eq:cancel}
	n&=3, \quad \frac{v}{\Mpl}=10^{-4}, \quad \kappa=-0.61, \quad \alpha=\frac{\varepsilon\Mpl^2}{v^2}=9.19, \nonumber \\ 
	g&=1.83\times10^{-3}, \quad c_\text{pot}=0.681, \quad c_\text{kin}=-0.676,
\end{align}
in Fig.~\ref{fig:scalar}.
The resultant PBH mass spectrum and induced GW spectrum are shown in Fig.~\ref{fig:PBH} and \ref{fig:GW}.
One can make another low-mass bump on the PBH spectrum, 
corresponding with the flat inflection point for $\kappa < 0$~\cite{Kawasaki:2016pql}. 
It may be detected by future gravitational lensing observations.
Also, the current PTA constraints are marginally avoided.

%%%%%%%%%%%%%%%%%%%%%%%%%%%%%%%%%
%%%%%%%%%%% Conclusions  %%%%%%%%%%%
%%%%%%%%%%%%%%%%%%%%%%%%%%%%%%%%%
\section{Conclusions}
\label{sec:conc}

In this paper, we have discussed the possibility to interpret the LIGO events
as mergers of PBHs that are produced by inflationary superhorizon fluctuations.
If scalar fluctuations during inflation are large enough, PBHs can be generated.
Interestingly,
while the formation of PBHs takes place at the horizon reentering of  overdense region,
GW is simultaneously produced through the second-order effects.
We have demonstrated that the PTA experiments together with the $\mu$ distortion of CMB play crucial roles
to probe/exclude this scenario.

The PTA experiments push the curvature perturbation to be steeper below the peak frequency
($\sim 3\times10^5\,\mathrm{Mpc}^{-1}$), while the $\mu$ distortion does above it.
Although the precise constraints depend on the required abundance of PBHs to account for the LIGO events, 
the spectral index has to be quite steep: $x \gtrsim 1.5$  and $y \gtrsim 2$ even for $\Omega_\text{PBH,tot} / \Omega_c \sim 10^{-4}$.
[Here we have used the parametrization of 
 $\mathcal P_\zeta$ given in Eq.~\eqref{eq:pzeta_prm}.]
The constraints become slightly stronger for a larger PBH fraction $\Omega_\text{PBH,tot} / \Omega_c \to 1$.
Note that simple small field inflation models are already disfavored because it is difficult to generate such a sharp spectrum owing to the slow roll condition.

There are several uncertainties on the obtained constraints, such as the uncertain factor $\gamma$, the threshold value of the density contrast $\delta_c$, and non-Gaussianity. 
First, recent numerical studies suggest a slightly larger value for the threshold value, $\delta_c\simeq0.45$~\cite{Musco:2004ak,Musco:2008hv,Musco:2012au}, 
than that of the simple analytical result, $\delta_c = 1/3$. Since the larger threshold results in stronger constraints, we take the latter one conservatively. 
Second, the viable range of $\gamma$ is restricted by the PTA experiments and the $\mu$ distortion as $0.02 \lesssim \gamma \lesssim 2$.
The representative value, $\gamma = 3^{-3/2}$, obtained from the simple analytic analysis is still allowed.
Finally, inflation models with an enhanced non-Gaussianity at the small scale can produce a larger/smaller amount of PBHs,
depending on the sign of the non-Gaussianity parameter.
Thus, the constraints on the curvature perturbation can be weaker/stronger for these models~\cite{Nakama:2016gzw}.

Then, we take the double inflation model proposed in Ref.~\cite{Kawasaki:1997ju} as an example.
Most of the parameters are already excluded because it is difficult to generate the required sharp spectrum due to the slow roll condition:
$\mathcal P_\zeta \propto k^{\ns - 1}$ 
with $\ns - 1 \simeq 2 \eta$ while $|\eta| \ll 1$.
We have proposed a new mechanism to obtain the sharp spectrum within this model. 
If the Hubble induced mass disappears after the end of chaotic inflation,
the scalar perturbation can be dramatically enhanced, which results in a steep GW spectrum.
In this case, the constraints can be marginally avoided.
Other than our scenario, such a sharp peak could be realized in \textit{e.g.}, multi-field double inflation~\cite{Frampton:2010sw,Kawasaki:2012kn},
single-field double inflation~\cite{Yokoyama:1998pt,Saito:2008em}, curvaton scenarios~\cite{Kawasaki:2012wr,Kohri:2012yw}, and gauge field production models~\cite{Garcia-Bellido:2016dkw,Cheng:2016qzb}.

Finally, let us summarize the future prospects of inflationary PBHs for the LIGO events.
The $\mu$ distortion as small as $\mu\sim10^{-9}$ may be probed by the future CMB observation, \textit{e.g.}, PIXIE~\cite{Kogut:2011xw} or PRISM~\cite{Andre:2013afa},
which could constrain the spectral index as large as $x\sim5$. Also the future PTA experiment, such as SKA, will probe the density parameter of GWS as small as $\Omega_\text{GW}h^2\sim10^{-15}$
and clarify/exclude the simple inflationary scenario which produce PBHs by Gaussian large scalar perturbations almost irrespective of the tilt $x$ and $y$.

%%%%%%%%%%%%%%%%%%%%%%%%%%%%%%%%%%
%%%%%%%%%%% Acknowledge %%%%%%%%%%%
%%%%%%%%%%%%%%%%%%%%%%%%%%%%%%%%%%
\section*{Acknowledgments}
{\small
\noindent
We would like to thank Ryo Saito and Jun'ichi Yokoyama for fruitful comments.
This work is supported by Grant-in-Aid for Scientific Research from the Ministry of Education,
Science, Sports, and Culture (MEXT), Japan,  No.\ 15H05889 (M.K.), No.\ 25400248 (M.K.),
No.\ 26104009 (T.T.Y.), No.\ 26287039 (T.T.Y.) and No.\ 16H02176 (T.T.Y.), 
World Premier International Research Center Initiative (WPI Initiative), MEXT, Japan 
(K.I., M.K., K.M., Y.T., and T.T.Y.),
JSPS Research Fellowships for Young Scientists (Y.T. and K.M.),
and Advanced Leading Graduate Course for Photon Science (K.I.).
}

\appendix
%%%%%%%%%%%%%%%%%%%%%%%%%%%%%%%%%%
%%%%%%%%%%% Appendix %%%%%%%%%%%
%%%%%%%%%%%%%%%%%%%%%%%%%%%%%%%%%%

\section{Derivation of GWs from Second-order Effects}
\label{app:gw_second}

\paragraph*{\bf Notation and conventions.}

Throughout this paper, we take the $(-+++)$ convention for the metric, $g_{\mu\nu}$.
The Riemann curvature tensor is defined as
\begin{align}
	R\indices{^\alpha_{\beta\mu\nu}} 
	= 2 \prn{
	\der\indices{_{[\mu}} \Gamma\indices{^\alpha_{\nu]\beta}} 
	+ \Gamma\indices{^\alpha_{\lambda [\mu}} \Gamma\indices{^\lambda_{\nu] \beta}}
	},
\end{align}
where the connection is given by
\begin{align}
	\Gamma\indices{^\alpha_{\beta \mu}} 
	= \frac{g\indices{^{\alpha \sigma}}}{2} 
	\prn{ 
	g\indices{_{\sigma \beta, \mu}} + g\indices{_{\mu \sigma, \beta}} -g\indices{_{\beta \mu, \sigma}}
	}.
	\label{eq:app_conn}
\end{align}
Here the square bracket $[\mu_1\mu_2 \cdots]$ represents anti-symmetrization of 
the tensor indices $\mu_1 \mu_2 \cdots$. The normalization of anti-symmetrization is defined as
\begin{align}
	[\mu_1 \mu_2 \cdots \mu_n ] = \frac{1}{n!} 
	\sum_{\sigma=(i_1,\dots, i_n)} \mathrm{sign} (\sigma)\,
	 \mu_{i_1}\mu_{i_2} \cdots \mu_{i_n}.
\end{align}
Here the summation is taken over all the permutations of $(i_1, \dots, i_n)$ denoted as $\sigma$.
For even (odd) permutations, the sign function, $\mathrm{sign} (\sigma)$, gives $+1$ ($-1$).
The Ricci tensor and the curvature scalar are defined respectively, by
\begin{align}
	R\indices{_{\mu\nu}} &= R\indices{^\alpha_{\mu\alpha\nu}},
	\label{eq:app_ricci} \\
	R &= R\indices{^\mu _\mu}.
	\label{eq:app_r}
\end{align}
The Einstein tensor is given by
\begin{align}
	G\indices{_{\mu\nu}} = R\indices{_{\mu\nu}} - \frac{g\indices{_{\mu\nu}}}{2} R.
\end{align}
Thus, the Einstein equation is
\begin{align}
	G\indices{_{\mu\nu}} = \frac{1}{\Mpl^2} T\indices{_{\mu\nu}},
\end{align}
with $T\indices{_{\mu\nu}}$ being the energy-momentum tensor.

In this paper, we consider the following perturbed metric:
\begin{align}
	g_{00} &= -a^2 ( 1 + 2 \Phi ), 	\label{eq:app_metric_00} \\
	g_{ij} &= a^2 \com{ (1 - 2 \Psi) \delta_{ij} + \frac{1}{2} h_{ij} }.
	\label{eq:app_metric_ij}
\end{align}
Here we take the Newton gauge and also neglect vector perturbations.
It involves scalar perturbations, $\Phi$ and $\Psi$,
and tensor perturbation, $h_{ij}$, which is the transverse-traceless component,
$\der_i h^{i}_j = h^i_i = 0$.
We assume that the lowest order tensor perturbation can be neglected
and consider that induced from second-order scalar perturbations.
Hence, we expect $h \sim \Phi^2 \sim \Psi^2$ for order counting.
The contravariant metric is obtained from $g_{\mu\nu}g^{\nu \sigma} = \delta_{\mu}^\sigma$:
\begin{align}
	g^{00} &= - \frac{1}{a^2} \prn{ 1 - 2 \Phi 
	}, \label{eq:app_contra00}\\
	g^{ij} &= \frac{1}{a^2} \com{ \prn{1 + 2 \Psi 
	} \delta^{ij} - \frac{1}{2} h^{ij}}. \label{eq:app_contraij}
\end{align}
Here we keep relevant terms sufficient for our purpose.
Note that the tensor perturbation with upper indices, $h^{ij}$, is defined by 
$h^{ij} = \delta^{ik} \delta^{jl} h_{kl}$

In this normalization of tensor perturbation,
the second-order action for tensor perturbation can be expressed as
\begin{align}
	 S_h =  \frac{\Mpl^2}{32} \int \dd \eta \dd^3 x\,  a^2
	 \com{ \prn{ h'_{ij}}^2 - \prn{ \nabla h_{ij} }^2 },
	\label{eq:app_tensor_action}
\end{align}
with $\eta$ being the conformal time, and the prime being a derivative with respect to $\eta$.
The energy density of graviton well inside the horizon may be given by
\begin{align}
	\rho_\text{GW} \simeq \frac{\Mpl^2}{16} \frac{1}{a^2} 
	\vev{ \overline{h_{ij, k} h_{ij,k}} },
	\label{eq:app_energy}
\end{align}
where the bracket indicates the expectation value,
and the overline stands for the oscillation average
to fulfill $\langle \overline{h'{}^2} \rangle \simeq \langle \overline{(\nabla h)^2} \rangle$.
It is useful to decompose $h_{ij}$ into two polarizations:
\begin{align}
	h_{ij} (\eta, \bm{x}) = 
	\int \frac{\dd^3 k}{(2 \pi)^{3/2}} \com{
		e_{ij} (\bm{k}) h_{\bm{k}} (\eta) + 	\bar e_{ij} (\bm{k}) \bar h_{\bm{k}} (\eta)
	} e^{i \bm{k} \cdot \bm{x}}.
	\label{eq:app_pol}
\end{align}
The polarization tensors are defined by
\begin{align}
	e_{ij} (\bm{k}) &\equiv 
	\frac{1}{\sqrt{2}} \com{ e_i (\bm{k}) e_j (\bm{k}) - \bar e_i (\bm{k}) \bar e_j (\bm{k}) }, 
	\label{eq:pol}\\
	\bar e_{ij} (\bm{k}) &\equiv 
	\frac{1}{\sqrt{2}} \com{ e_i (\bm{k}) \bar e_j (\bm{k}) + \bar e_i (\bm{k}) e_j (\bm{k}) },
	\label{eq:pol_bar}
\end{align}
where $\bm{e}(\bm{k})$ and $\bar{\bm{e}}(\bm{k})$ are 
two independent unit vectors orthogonal to $\bm{k}$, satisfying $\bm e \cdot \bar{\bm e} = 0$.
Thus, the polarization tensors satisfy the following relations:
$e_{ij} e_{ij} = \bar e_{ij} \bar e_{ij} = 1$ and $e_{ij} \bar e_{ij} = 0$.
Assuming the translational invariance, let us define the power spectrum of tensor perturbation
\begin{align}
	\vev{h_{\bm{k}} (\eta) h_q (\eta)} =: \delta (\bm{k} + \bm{q}) \times \frac{2 \pi^2}{k^3}
	\mathcal P_h (\eta,k).
	\label{eq:app_power}
\end{align}
We expect $\mathcal P_h = \mathcal P_{\bar h}$ for $CP$ conserving background.

Plugging Eq.~\eqref{eq:app_pol} into Eq.~\eqref{eq:app_energy}, and using Eq.~\eqref{eq:app_power},
one can express the energy density of GWs as a function of tensor power spectrum:
\begin{align}
	\rho_\text{GW} (\eta) &=\int \dd \ln k \, \rho_\text{GW} (\eta, k), \\
	\rho_\text{GW} (\eta,k) &\equiv \frac{\Mpl^2}{8}  \prn{\frac{k}{a}}^2 
	\overline{ \mathcal P_h (\eta, k) }.
	\label{eq:app_energy_power}
\end{align}
Here we have used the fact that the polarization tensors defined in Eqs.~\eqref{eq:pol} and \eqref{eq:pol_bar}
are normalized by $e_{ij} e_{ij} = \bar e_{ij} \bar e_{ij} = 1$.
It is useful to define the density parameter of GWs normalized by the critical density of the Universe:
\begin{align}
	\Omega_\text{GW} (\eta, k) \equiv \frac{\rho_\text{GW} (\eta, k)}{\rho_\text{crit}}
	= \frac{1}{24} \prn{ \frac{k}{a H}}^2 \overline{ \mathcal P_h (\eta, k) }.
	\label{eq:app_Omega_gw}
\end{align}
\paragraph*{\bf Second-order Einstein equation.}

Here we summarize the second-order Einstein equation which is required to study 
GWs production from the second-order effects.
For clarity, we summarize basic assumptions (some of them are already mentioned):
(i) negligible vector perturbations, (ii) negligible first order tensor perturbation,
and (iii) negligible skewness, \textit{i.e.,}~$\Phi = \Psi$.

The spatial component of second-order Einstein tensor is given by
\begin{align}
	\hat{\mathcal T}\indices{_{ij;kl}} \tensor{\delta}*{_{kk'}} G\indices{^{(2)k'}_{l}} 
	&= \frac{1}{a^2} \com{
		\frac{1}{4} \prn{ h''_{ij} + 2 \mathcal H h'_{ij} - \nabla^2 h_{ij} }
		+ \hat{\mathcal T}_{ij;kl} \tilde S_{kl}
	},
\end{align}
where
\begin{align}
	\tilde S_{ij} = 4 \Psi \der_i \der_j \Psi + 2 \der_i \Psi \der_j \Psi.
\end{align}
The second-order energy-momentum tensor is
\begin{align}
		&\hat{\mathcal T}\indices{_{ij;kl}} \tensor{\delta}*{_{kk'}} T\indices{^{(2)k'}_{l}} \nonumber \\
		&= 
		\hat{\mathcal T}\indices{_{ij;kl}}
		\frac{\Mpl^2}{a^2} \com{
			\frac{4}{3 (1 + w)} \der_k \prn{ \frac{\Psi'}{\mathcal H} + \Psi} \der_l \prn{\frac{\Psi'}{\mathcal H} + \Psi}
		}.
\end{align}
To extract the transverse-traceless component,
we have used the projection operator, $\mathcal P_{ij;kl}$.
It acts as a projection to the transverse-traceless part:
\begin{align}
	\hat{\mathcal T}_{ij;kl} \hat{\mathcal T}_{kl;mn} = \hat{\mathcal T}_{ij;mn}, \quad
	\der_i \hat{\mathcal T}_{ij;kl}\, \bullet_{kl} = \hat{\mathcal T}_{ii;kl}\, \bullet_{kl} = 0,
	\label{eq:app_tensor_proj}
\end{align}
where $\bullet_{ij}$ is an arbitrary second-order tensor.
From the transverse-traceless component of the second-order Einstein equation
\begin{align}
	\hat{\mathcal T}\indices{_{ij;kl}} \tensor{\delta}*{_{kk'}} G\indices{^{(2)k'}_{l}} 
	= \frac{1}{\Mpl^2} \hat{\mathcal T}\indices{_{ij;kl}} \tensor{\delta}*{_{kk'}} T\indices{^{(2)k'}_{l}},
\end{align}
we obtain the equation of motion for GWs
\begin{align}
	h''_{ij} + 2 \mathcal H h'_{ij} - \nabla^2 h_{ij} 
	= - 4\hat{\mathcal T}_{ij;kl} S_{kl},
\end{align}
where the source term is given by
\begin{align}
	S_{ij} \equiv& 4 \Psi \der_i \der_j \Psi + 2 \der_i \Psi \der_j \Psi \\
	 &- 
	\frac{4}{3 (1 + w)} \der_i \prn{ \frac{\Psi'}{\mathcal H} + \Psi} \der_j \prn{\frac{\Psi'}{\mathcal H} + \Psi}.
\end{align}
Here we have used $\hat{\mathcal T}_{ij;kl} h_{kl} = h_{ij}$.
In terms of polarization tensors,
the source term can be expressed as 
\begin{align}
	\hat{\mathcal{T}}_{ij;kl} S_{kl} = \int \frac{\dd^3 k}{(2\pi)^{3/2}} e^{i \bm{k} \cdot \bm{x}}
	\hat{\mathcal T}_{ij;kl} ( \bm{k})  S_{kl} (\bm{k}),
\end{align}
where
\begin{align}
	\hat{\mathcal T}_{ij;kl} ( \bm{k}) \equiv e_{ij}( \bm{k}) e_{kl}( \bm{k}) + \bar e_{ij}( \bm{k}) \bar e_{kl}( \bm{k}).
\end{align}
One can see that $\hat{\mathcal T}_{ij;kl} (\bm{k})$ satisfies Eq.~\eqref{eq:app_tensor_proj}.
By performing the Fourier transform,
one obtains the equation of motion for each polarization of GWs as follows:
\begin{align}
	h_{\bm{k}}'' (\eta) + 2 \mathcal H h'_k (\eta) + k^2 h_{\bm{k}} (\eta) = 4S_{\bm{k}} (\eta),
	\label{eq:app_eom_pol}
\end{align}
where
\begin{align}
	\begin{split}
	S_{\bm{k}} (\eta) = \int & \frac{\dd^3 q}{(2\pi)^{3/2}} e_{ij}(\bm{k}) q_i q_j\, 
		\Bigg[ 2\Psi_{\bm{q}} (\eta) \Psi_{\bm{k}-\bm{q}} (\eta)  \\[.5em]
		&+ 
		\prn{ \frac{\Psi'_{\bm{q}} (\eta)}{\mathcal H} +\Psi_{\bm{q}} (\eta) } 
		\prn{  \frac{\Psi'_{\bm{k}-\bm{q}} (\eta)}{\mathcal H} +\Psi_{\bm{k}-\bm{q}} (\eta)  } 
		\Bigg].
	\end{split}
\end{align}
Here the Fourier transform of scalar perturbation is defined by
\begin{align}
 	\Psi (\eta, \bm{x}) = \int \frac{\dd^3 k}{(2\pi)^{3/2}} e^{i \bm{k} \cdot \bm{x}} \Psi_{\bm{k}} (\eta).
\end{align}
\paragraph*{\bf Energy density of GWs.}
Here we derive a useful expression for the energy density of GWs
using Eqs.~\eqref{eq:app_power} and \eqref{eq:app_Omega_gw}.
The equation of motion for $h_{\bm{k}}$, given in Eq.~\eqref{eq:app_eom_pol}, 
can formally be solved by means of the Green function:
\begin{align}
	h_{\bm{k}} (\eta) 
	= \frac{4}{a (\eta)} \int^\eta \dd \eta' G_{\bm{k}}^{(h)} (\eta, \eta') 
	\com{ a (\eta') S_{\bm{k}} (\eta') },
	\label{eq:app_formal_sol}
\end{align}
where 
\begin{align}
	G^{(h)''}_{\bm{k}} (\eta, \eta') + \prn{ k^2 - \frac{a''}{a} } G^{(h)}_{\bm{k}} (\eta , \eta') = \delta (\eta - \eta').
\end{align}
We are mostly interested in scales which enter the horizon in the radiation-dominated era,
and we also assume that the Universe is dominated by radiation
from the reheating to the matter-radiation equality, $\eta_\text{eq}$.
Hence, the Green function during the radiation-dominated era is sufficient for our purpose:
\begin{align}
	G^{(h)}_{\bm{k}} (\eta , \eta') =  \frac{\sin \com{ k \prn{\eta - \eta' }}}{k} ~~ \text{for} ~~ \eta > \eta'.
\end{align}

The linear order scalar perturbation can be split into two parts;
the primordial perturbation, $\psi_{\bm{k}}$, and the transfer function, $\Psi (k \eta)$:
\begin{align}
	\Psi_{\bm{k}} (\eta) = \Psi (x) \psi_{\bm{k}} ~~
	\text{with} ~~ x \equiv k \eta.
\end{align}
The explicit form of the transfer function during the radiation-dominated era is given by
\begin{align}
	\Psi (x) = \frac{9}{x^2} \com{ \frac{\sin \prn{x/\sqrt{3}}}{x/\sqrt{3}} - \cos \prn{x/\sqrt{3}}}~~
	\text{for} ~~ \eta < \eta_\text{eq}.
	\label{eq:app_scalar transfer}
\end{align}
In the matter-dominated era, the transfer function becomes constant, $\Psi (x) = \text{const.}$.
Assuming the translational invariance,
we define the power spectrum of the primordial scalar perturbation as follows:
\begin{align}
	\vev{ \psi_{\bm{k}} \psi_{\bm{q}}} \equiv \delta (\bm{k} + \bm{q}) \times \frac{2 \pi^2}{k^3}
	\mathcal P_\Psi (k).
\end{align}
The scalar perturbation, $\Psi$, is related 
to the curvature perturbation, $\zeta$, for $k \eta \ll 1$
as $\Psi = - 2 \zeta / 3$ in the radiation-dominated era,
which indicates
\begin{align}
	\mathcal P_\Psi (k) = \frac{4}{9} \mathcal P_\zeta (k).
	\label{eq:app_conversion}
\end{align}
%%
\begin{comment}
Note that we are interested in modes that enter the horizon during the radiation-dominated era,
and that the conserved quantity on superhorizon scales is the curvature perturbation.
For these reasons, we have to use the relation Eq.~\eqref{eq:app_conversion} 
in order to estimate the primordial scalar power spectrum, $\mathcal P_\Psi$.
\end{comment}

Now we are in a position to derive a formula which relates the curvature power spectrum to the tensor power spectrum.
As can be seen from Eq.~\eqref{eq:app_formal_sol},
the power spectrum of GWs can be obtained from a two point function of the source term:
\begin{align}
	\begin{split}
	\vev{ S_{\bm{k}} (\eta_1) S_{\bm{k}'} (\eta_2)}
	=
	\int & \frac{\dd^3 \tilde k \dd^3 \tilde k'}{(2 \pi)^{3}}
	e_{ij} (\bm{k}) \tilde k_i \tilde k_j e_{ij} (\bm{k}') \tilde k'_i \tilde k'_j  
	\\ 
	\times f (\tilde{\bm{k}}, \bm{k} - \tilde{\bm{k}}, \eta_1)
	&f (\tilde{\bm{k}}', \bm{k}' - \tilde{\bm{k}}', \eta_2) 
	\vev{
		\psi_{\bm{k} - \tilde{\bm{k}}} \psi_{\bm{k}} \psi_{\bm{k}' - \tilde{\bm{k}}'} \psi_{\bm{k}'} 
		},
	\end{split}
\end{align}
where
\begin{align}
	f(\bm{k_1}, \bm{k_2}, \eta)  \equiv & \,
	2\Psi (x_1) \Psi  (x_2) \nonumber  \\
		&+ 
		\prn{ \frac{\Psi' (x_1)}{\mathcal H} +\Psi (x_1) } 
		\prn{  \frac{\Psi'(x_2)}{\mathcal H} +\Psi (x_2)  },
\end{align}
with $x_i \equiv k_i \eta$.	
Here note that the prime stands for a derivative with respect to $\eta$ not $x$.
One can clearly see that $f(\bm{k_1}, \bm{k_2}, \eta)$ is symmetric 
under $\bm{k_1} \leftrightarrow \bm{k_2}$, and that it is invariant under $\bm{k_i} \to - \bm{k_i}$.
For our normalization of the polarization tensors, given in Eq.~\eqref{eq:pol} [Eq.~\eqref{eq:pol_bar}],
the projection becomes $e_{ij} (\bm{k}) \tilde k_i \tilde k_j = \tilde k^2 \sin^2 \theta \cos 2 \varphi / \sqrt{2}$
with $\theta$ and $\varphi$ being defined by $\cos \theta \equiv \bm{k} \cdot \tilde{\bm{k}}/k \tilde k$ 
and by the azimuthal angle of $\tilde{\bm{k}}$, respectively.
Non-vanishing contributions of 
$\vev{\psi_{\bm{k} - \tilde{\bm{k}}} \psi_{\bm{k}} \psi_{\bm{k}' - \tilde{\bm{k}}'} \psi_{\bm{k}'}}$
come from two connected parts, which are proportional to
$\delta (\bm{k} + \bm{k}' ) \delta ( \tilde{\bm{k}} + \tilde{\bm{k}}' ) 
\mathcal P_\Psi (\tilde k) \mathcal P_\Psi (| \bm{k} - \tilde{\bm{k}} |)$
or
$\delta (\bm{k} + \bm{k}' ) \delta ( \bm{k} - \tilde{\bm{k}} + \tilde{\bm{k}}' ) 
\mathcal P_\Psi (\tilde k) \mathcal P_\Psi (| \bm{k} - \tilde{\bm{k}} |)$.
Noting that $f (\bm{k_1}, \bm{k_2}, \eta)$ is invariant under $\bm{k_1} \leftrightarrow \bm{k_2}$ 
and/or $\bm{k_i} \to - \bm{k_i}$,
and that the projection $e_{ij} (\bm{k}) \tilde k_i \tilde k_j$ is invariant under
$\bm{k} \to - \bm{k}$ and/or $\tilde{\bm{k}} \to - \tilde{\bm{k}}$,
one can see that these two contributions are essentially the same
after the integration of $\tilde{\bm{k}}'$.
Thus, one finds
\begin{align}
	& \vev{ S_{\bm{k}} (\eta_1) S_{\bm{k}'} (\eta_2)} \nonumber\\
	& = 
	\frac{2 \pi^2}{k^3} \delta (\bm{k} + \bm{k}') 
	\frac{4}{81}\int_0^\infty \dd\tilde k \int_{-1}^1 \dd \mu \, \frac{k^3 \tilde k^3}{\abs{\bm{k} - \tilde{\bm{k}}}^3} \prn{ 1 - \mu^2 }^2 \nonumber\\
	&\qquad \times f (\tilde{\bm{k}}, \bm{k} - \tilde{\bm{k}}, \eta_1) f (\tilde{\bm{k}}, \bm{k} - \tilde{\bm{k}}, \eta_2)
	\mathcal P_\zeta (\tilde k) \mathcal P_\zeta (| \bm{k} - \tilde{\bm{k}}|).
\end{align}
To simplify the expression, let us rewrite the equation in terms of 
$u \equiv |\bm{k} - \tilde{\bm{k}}| / k$ and $v \equiv \tilde k / k$ 
instead of $\tilde k$ and $\mu$.
After some algebra, we eventually get the following formula for the tensor power spectrum:
\begin{align}
	\mathcal P_h (\eta,k) = &\, \frac{64}{81} \prn{\frac{a_0 /\eta_0}{k a (\eta)}}^2
	\int_0^\infty \dd v \int_{|1- v|}^{1+v} \dd u \,  I^2 (v, u, x) \nonumber \\
	&\times \com{ \frac{4 v^2 - \prn{ 1- u^2 + v^2 }^2 }{4 v u} }^2 
	\mathcal P_\zeta (k v) \mathcal P_\zeta (k u),
\label{eq:app_power}
\end{align}
where 
\begin{align}
	I (v,u,x) &\equiv \int^x_0 \dd \bar x \com{k \frac{a(\bar \eta) \eta_0}{a_0} } \com{k G_{\bm{k}} (\eta , \bar \eta)} 
	f (\tilde{\bm{k}}, \bm{k} - \tilde{\bm{k}}, \bar \eta) \nonumber \\
	\begin{split}
	&= \int^x_0 \dd \bar x \bar x \sin \prn{ x - \bar x } \Big[  
		3 \Psi (v \bar x) \Psi (u \bar x) \\
		& \qquad \quad + \bar x \big\{ \Psi (v \bar x) u \Psi' (u \bar x) + v \Psi' (v \bar x) \Psi (u \bar x)\big\} \\
		& \qquad \qquad \qquad \qquad \qquad \quad + \bar x^2 u v \Psi' (u \bar x) \Psi' (v \bar x)  \Big],
	\end{split}
	\label{eq:app_Ivux}
\end{align} 
with $x \equiv k \eta$, and $\bar x \equiv k \bar \eta$.
Note here that $\Psi'$ does not represent the derivative of $\Psi$ with respect to $\eta$, 
but instead, with respect to $x$.
This expression is valid for modes that enter the horizon in the radiation-dominated era,
as long as $\eta < \eta_\text{eq}$.

Finally, let us discuss the current energy density of GWs.
The density parameter of GWs remains almost constant for a sufficiently large $\eta$
during the radiation-dominated era, \textit{i.e.,}~$\eta < \eta_\text{eq}$,
because the growth of the scale factor $a^4 (\eta)$ and the decay of $H^2$ cancel out
[see Eqs.~\eqref{eq:app_Omega_gw} and \eqref{eq:app_power}].
After the matter-radiation equality, $\eta > \eta_\text{eq}$,
the amount of energy carried by GWs decreases because GWs well inside the horizon
behave as radiation while the Universe is dominated by matter.
Multiplying this dilution factor by the cosmic expansion,
we obtain the following expression for the current density parameter of GWs:
\begin{align}
	\Omega_\text{GW} (\eta_0, k) 
	= \prn{ \frac{a_\text{c}^2 H_\text{c}}{a_0^2 H_0} }^2 \Omega_\text{GW} (\eta_\text{c}, k)
	=  \Omega_{r,0} \Omega_\text{GW} (\eta_c, k),
	\label{eq:app_omega_current}
\end{align}
where $a_0 (=1)$ and $a_c$ are scale factors 
at present and at the time $\Omega_\text{GW}$ becomes constant respectively.
$\eta_c (< \eta_\text{eq})$ represents a time before the matter-radiation equality
after which the GW density parameter becomes constant.
$\Omega_{r,0}$ is the density parameter of radiation at present.
Thus, all we have to compute is the density parameter at $\eta_c$:
\begin{align}
\begin{split}
	\Omega_\text{GW} (\eta_c, k) 
	= \frac{8}{243} \int_0^\infty \dd v \int_{|1 - v|}^{1 + v} \dd u 
	\com{ \frac{4 v^2 - \prn{ 1- u^2 + v^2 }^2 }{4 v u} }^2 &\\
	\quad \times \mathcal P_\zeta (k v) \mathcal P_\zeta (k u) \overline{ I^2 (v, u, k / k_c) }.&
\end{split}
\label{eq:app_omega_eq}
\end{align}
Here we have used the useful formulas valid in the radiation-dominated era:
$a = a_0 \eta / \eta_0$, $\mathcal H = a H = \eta^{-1}$ 
and $x_c = k \eta_c = k / k_c$
with $k_c$ being a comoving momentum that enters the horizon at $\eta_c$.

\begin{comment}
It is instructive to consider the flat power spectrum for the curvature perturbation,
$\mathcal P_\zeta = A$, to obtain an order of magnitude estimation
of $\Omega_\text{GW}$. We find
%%
\begin{align}
	\Omega_\text{GW}h^2 \sim 10^{-9} \, \prn{ \frac{A}{0.01} }^2.
	\label{eq:app_omega_gw_order}
\end{align}
%%
One can see that large curvature perturbation required for PBH formation, 
$\mathcal P_\zeta \sim \mathcal O (0.01)$, yields a substantial amount of GWs.
\end{comment}

Finally, we would like to comment on one missing effect in this calculation.
As emphasized in Refs.~\cite{Baumann:2007zm,Bugaev:2009zh}, this simple argument misses the fact that
the GWs in the subhorizon regime do not always propagate freely because of the presence
of the source term. Before the matter-radiation equality, the source term decreases faster than the GWs amplitude,
and thus we can safely neglect this effect. But, afterwards, the source term becomes constant
while the amplitude of GWs decreases by the cosmic expansion.
As a result, GWs produced by this constant source term could dominate over those produced at the horizon reentry
in the radiation-dominated era.
As discussed in Refs.~\cite{Baumann:2007zm,Bugaev:2009zh}, this effect can be neglected for a sufficiently small scale 
$k \gtrsim  k_\text{crit} \simeq 1 \, \mathrm{Mpc}^{-1}$.
Since we are interested in scales shorter than this scale, we can rely on the simply formula given 
in Eqs.~\eqref{eq:app_omega_current} and \eqref{eq:app_omega_eq}.

\smallskip
\paragraph*{\bf Monochromatic scalar perturbation.}

As the end of this appendix, let us see the analytic approximation of GWs in the case of the monochromatic scalar perturbations, that is,
\bae{\label{eq:app_deltapower}
	\calP_\zeta(k)=A_\zeta\delta(\log k-\log k_*).
}
For this power spectrum, the GW density parameter~(\ref{eq:app_omega_eq}) reads
\bae{\label{eq:app_omega_delta}
	\Omega_\text{GW}(\eta_c,k)=&\frac{8}{243}\theta\left(1-\frac{k}{2k_*}\right)\left[1-\left(\frac{k}{2k_*}\right)^2\right]^2\left(\frac{k_*}{k}\right)^2 \nonumber \\
	&\times\overline{I^2\left(\frac{k_*}{k},\frac{k_*}{k},\frac{k}{k_c}\right)}\,A_\zeta^2,
}
where $\theta(x)$ is the step function. $I(k_*/k,k_*/k,k/k_c)$ is given by Eq.~(\ref{eq:app_Ivux}) as
\bae{
	&I\left(\frac{k_*}{k},\frac{k_*}{k},\frac{k}{k_c}\right)=\int^{k/k_c}_0\dd\bar{x}\sin\left(\frac{k}{k_c}-\bar{x}\right)\bar{x} \nonumber \\
	&\times\left[3\Psi^2(k_*\bar{\eta})+2k_*\bar{\eta}\Psi(k_*\bar{\eta})\Psi^\prime(k_*\bar{\eta})+k_*^2\bar{\eta}^2\Psi^\prime(k_*\bar{\eta})\right].
}
With use of the dominant terms of the integrant for large $\bar{\eta}$ for simplicity, it can be approximated as
\bae{
	&I\left(\frac{k_*}{k},\frac{k_*}{k},\frac{k}{k_c}\right) \nonumber \\
	&\sim\frac{27}{2}\left(\frac{k}{k_*}\right)^2\int^{k_*/k_c}_0\dd\bar{x}_*\sin\left(\frac{k}{k_c}-\frac{k}{k_*}\bar{x}_*\right)
	\frac{1}{\bar{x}_*}\left[1-\cos\left(\frac{2}{\sqrt{3}}\bar{x}_*\right)\right],
}
after changing the integration variable to $\bar{x}_*=k_*\bar{\eta}$.
This integral actually has a logarithmic singularity at $k=2k_*/\sqrt{3}$; otherwise, it can be solved as
\begin{widetext}
\bae{
	I\left(\frac{k_*}{k},\frac{k_*}{k},\frac{k}{k_c}\right)\sim&\frac{27}{4}\left(\frac{k}{k_*}\right)^2\left[\left(\Si\left(\left(\frac{2}{\sqrt{3}}+\frac{k}{k_*}\right)\frac{k_*}{k_c}\right)-2\Si\left(\frac{k}{k_c}\right)
	-\Si\left(\left(\frac{2}{\sqrt{3}}-\frac{k}{k_*}\right)\frac{k_*}{k_c}\right)\right)\cos\left(\frac{k}{k_c}\right)\right. \nonumber \\
	&\left.-\left(\Ci\left(\left(\frac{2}{\sqrt{3}}+\frac{k}{k_*}\right)\frac{k_*}{k_c}\right)-2\Ci\left(\frac{k}{k_c}\right)
	+\Ci\left(\left|\frac{2}{\sqrt{3}}-\frac{k}{k_*}\right|\frac{k_*}{k_c}\right)-\log\left|\frac{4}{3}\frac{k_*^2}{k^2}-1\right|\right)\sin\left(\frac{k}{k_c}\right)\right].
}
\end{widetext}
Here $\Si(x)$ and $\Ci(x)$ are the sine integral and the cosine integral defined by
\bae{
	\Si(x)=\int^x_0\frac{\sin t}{t}\dd t, \quad\quad \Ci(x)=-\int^\infty_x\frac{\cos t}{t}\dd t.
}
With use of their limits:
\bae{
	\lim_{x\to\pm\infty}\Si(x)=\pm\frac{\pi}{2}, \quad\quad \lim_{x\to\infty}\Ci(x)=0,
}
the asymptotic form of $I(k_*/k,k_*/k,k/k_c)$ for large $\eta_c=k_c^{-1}$ reads
\begin{widetext}
\bae{
	I\left(\frac{k_*}{k},\frac{k_*}{k},\frac{k}{k_c}\right)\,\overset{\eta_c\to\infty}{\sim}
	\bce{
		\displaystyle
		-\frac{27}{4}\left(\frac{k}{k_*}\right)^2\left[\pi\cos\left(\frac{k}{k_c}\right)-\log\left(\frac{4}{3}\frac{k_*^2}{k^2}-1\right)\sin\left(\frac{k}{k_c}\right)\right], 
		& \displaystyle
		\text{for}\,\,\, k<\frac{2}{\sqrt{3}}k_*, \\[10pt]
		\displaystyle
		\frac{27}{4}\left(\frac{k}{k_*}\right)^2\log\left(1-\frac{4}{3}\frac{k_*^2}{k^2}\right)\sin\left(\frac{k}{k_c}\right), 
		& \displaystyle
		\text{for}\,\,\, k>\frac{2}{\sqrt{3}}k_*.
	}
}
Therefore one can obtain its squared $k\eta_c$-periodic average as
\bae{\label{eq:app_I2bar}
	\overline{I^2\left(\frac{k_*}{k},\frac{k_*}{k},\frac{k}{k_c}\right)}\sim
	\bce{
		\displaystyle
		\frac{1}{2}\left(\frac{27}{4}\right)^2\left(\frac{k}{k_*}\right)^4\left(\pi^2+\log^2\left(\frac{4}{3}\frac{k_*^2}{k^2}-1\right)\right),
		& \displaystyle
		\text{for}\,\,\, k<\frac{2}{\sqrt{3}}k_*, \\[10pt]
		\displaystyle
		\frac{1}{2}\left(\frac{27}{4}\right)^2\left(\frac{k}{k_*}\right)^4\log^2\left(1-\frac{4}{3}\frac{k_*^2}{k^2}\right),
		& \displaystyle
		\text{for}\,\,\, k>\frac{2}{\sqrt{3}}k_*,
	}
}
\end{widetext}
after a composition of the trigonometric functions. Substituting it into Eqs.~(\ref{eq:app_omega_current}) and (\ref{eq:app_omega_delta}),
one can see the current density parameter of GWs induced by the monochromatic scalar perturbations~(\ref{eq:app_deltapower}).

Eq.~(\ref{eq:app_I2bar}) obviously shows a singular peak at $k=2k_*/\sqrt{3}$, but the integral around this singularity does converge.
For example, the current GW density for one logarithmic wavelength bin around the peak scale $k_p=2k_*/\sqrt{3}$ can be evaluated as
\bae{\label{eq:app_peakGW}
	&\int^{\log k_p+1/2}_{\log k_p-1/2}\dd\log k\,\Omega_\text{GW}(\eta_0,k)h^2 \nonumber \\
	&\quad\quad\simeq1.2\times10^{-8}\left(\frac{\Omega_{r,0}h^2}{4.2\times10^{-5}}\right)\left(\frac{A_\zeta}{0.01}\right)^2.
}
Therefore  one can use this value practically for the peak GW.

\section{Analytic comprehension for multiple horizon crossing modes}
\label{sec:multihorizon}

Here we proceed with the analytic comprehension for the modes which exit the horizon around the end of the pre-inflation,
reenter the horizon during the oscillation phase, and then reexit the horizon around the beginning of the second new inflation,
extending the discussion in the appendix of Ref.~\cite{Kawasaki:2016pql}.

The linear equation of motion (EoM) for perturbations which have a generic Hubble induced mass term $\frac{3}{2}cH^2\varphi^2$ is given by
\begin{align}
	0\sim\begin{cases}
		\displaystyle
		\delta\ddot{\varphi}+3H\delta\dot{\varphi}+3cH^2\delta\varphi, & k\ll aH, \\[5pt]
		\displaystyle
		\delta\ddot{\varphi}+3H\delta\dot{\varphi}+\frac{k^2}{a^2}\delta\varphi. & k\gg aH.
	\end{cases}
\end{align}
Here we neglected the effect of metric perturbations for simplicity though we include them in the main body of this paper.

The subhorizon EoM can be rewritten as
\begin{align}
	\partial_\eta^2(a\delta\varphi)+k^2(a\delta\varphi)\simeq0,
\end{align}
with use of the conformal time $a\dd\eta=\dd t$ and in the subhorizon limit.
Therefore it only has oscillating solutions whose amplitudes decrease as $a^{-1}$.
On the other hand, in the case where the background equation of state is given by $w=p/\rho>-1\,(a\propto t^\frac{2}{3(1+w)})$,
the superhorizon EoM reads
\begin{align}
	\delta\ddot{\varphi}+\frac{2}{(1+w)t}\delta\dot{\varphi}+\frac{4c}{3(1+w)^2t^2}\delta\varphi=0.
\end{align}
Its dominant mode decays as $t^{-\mathrm{Re}\left[\frac{1-w}{2(1+w)}\left(1-\sqrt{1-\frac{16c}{3(1-w)^2}}\right)\right]}$,
that is, the decay factor is $t^{-\frac{1-w}{2(1+w)}}\propto a^{-\frac{3(1-w)}{4}}$ if $c\geq\frac{3(1-w)^2}{16}$,
and otherwise $t^{-\frac{1-w}{2(1+w)}\left(1-\sqrt{1-\frac{16c}{3(1-w)^2}}\right)}\propto a^{-\frac{3(1-w)}{4}\left(1-\sqrt{1-\frac{16c}{3(1-w)^2}}\right)}$.
In particular, in the massless limit $c\to0$, the dominant mode is constant consistently with intuition.
In the exact de Sitter background, the two solutions can easily be found as 
$\delta\varphi\propto \exp\left[-\frac{3}{2}Ht\left(1\pm\sqrt{1-\frac{4}{3}c}\right)\right]$ since the Hubble parameter is constant.
Therefore the dominant mode damps as $a^{-\frac{3}{2}\left(1-\mathrm{Re}\sqrt{1-\frac{4}{3}c}\right)}$ and actually this damping factor
is an extension of the above one for $w>-1$ to the de Sitter case $w=-1$.
In summary, the amplitude of the perturbations decreases as $a^{-1}$ in the subhorizon limit and as 
$a^{-\frac{3(1-w)}{4}+\mathrm{Re}\nu}$ in the superhorizon limit where
$\nu$ is defined as $\nu=\sqrt{\left(\frac{3(1-w)}{4}\right)^2-3c}$.

\begin{figure}
	\centering
	\includegraphics[width=0.9\hsize]{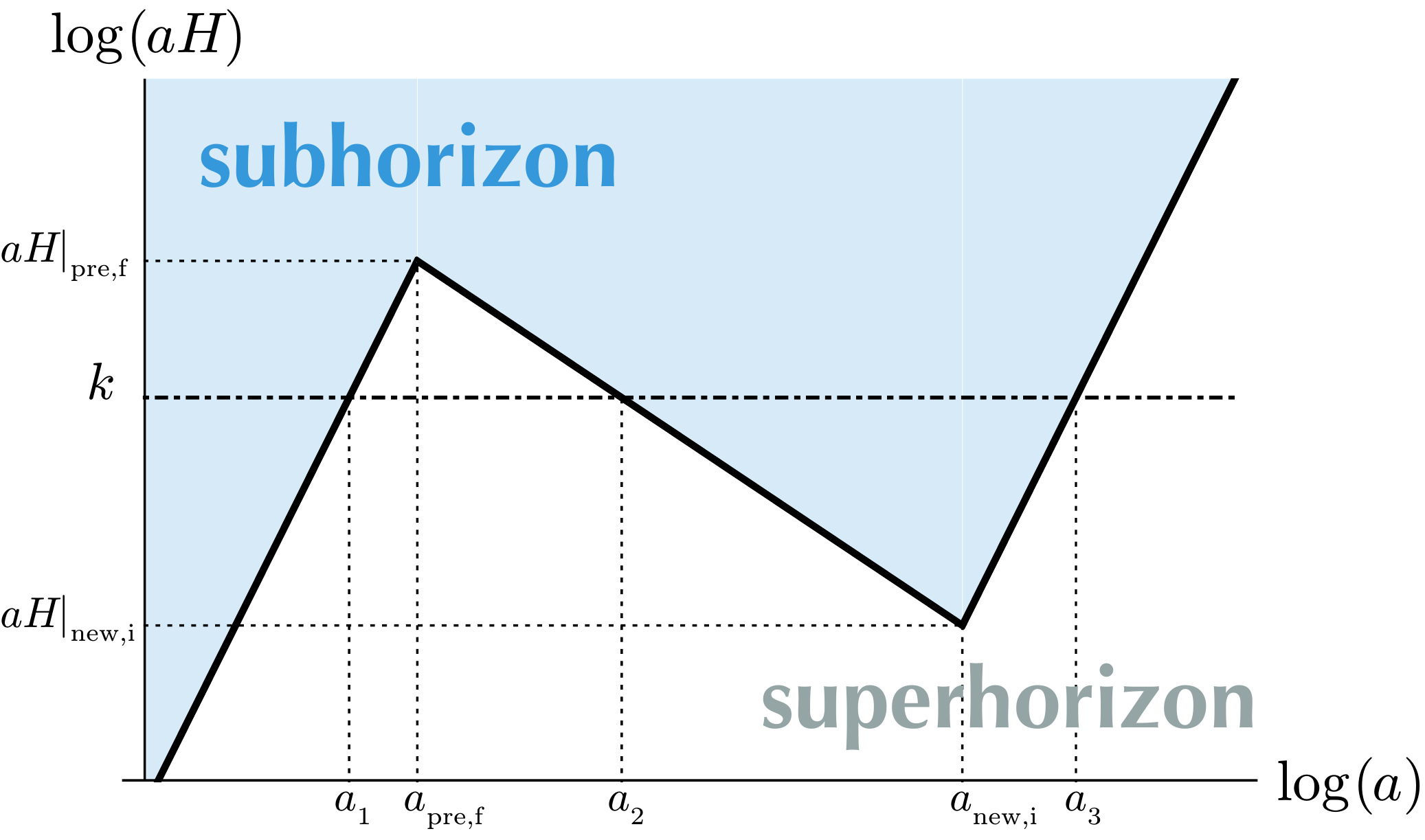}
	\caption{The schematic image of the multiple horizon crossing mode $k$.
	$a_\text{pre,f}$ and $a_\text{new,i}$ represent the time of the end of the pre-inflation and the beginning of the second new inflation.}
	\label{fig:aH}
\end{figure}

Now let us evaluate the total damping factor, following the schematic illustration of Ref.~\cite{Kawasaki:2016pql} as Fig.~\ref{fig:aH}.
In this image, $a_\text{pre,f}$ and $a_\text{new,i}$ represent the time of the end of the pre-inflation and the beginning of the second new inflation,
and the mode $k$ crosses the horizon three times, at $a_1$, $a_2$, and $a_3$.
Before calculating the total damping factor, note that the coefficient of the Hubble induced mass $c$ during the pre-inflation can be
different from that during the oscillation phase. In terms of the notation of the main text, the former one is given by $c_\text{pot}$
and the latter one is $c=\frac{1}{2}(c_\text{pot}+c_\text{kin})$ if $\phi$ oscillates with a quadratic potential.
Let us refer to their corresponding $\nu$ as $\nu_\text{pre}$ and $\nu_\text{osc}$ respectively.

Then, with use of the above decaying formula, the amplitude of $\delta\varphi$ at $a_3$ is estimated as
\begin{align}
	\delta\varphi|_3\sim&\,\frac{H_\text{pre}}{2\pi}\left(\frac{a_\text{pre,f}}{a_1}\right)^{-\frac{3}{2}+\mathrm{Re}\nu_\text{pre}}
	\left(\frac{a_2}{a_\text{pre,f}}\right)^{-\frac{3(1-w)}{4}+\mathrm{Re}\nu_\text{osc}} \nonumber \\
	&\times\left(\frac{a_\text{new,i}}{a_2}\right)^{-1}\left(\frac{a_3}{a_\text{new,i}}\right)^{-1}.
\end{align}
Noting that the horizon scale $aH$ is proportional to $a^{-\frac{1+3w}{2}}$, it can be rewritten as
\begin{align}
	\delta\varphi|_3\sim&\,\frac{H_\text{pre}}{2\pi}\left(\frac{aH|_\text{pre,f}}{k}\right)^{-\frac{3}{2}+\mathrm{Re}\nu_\text{pre}}
	\left(\frac{k}{aH|_\text{pre,f}}\right)^{\frac{3(1-w)}{2(1+3w)}-\frac{2}{1+3w}\mathrm{Re}\nu_\text{osc}} \nonumber \\
	&\times\left(\frac{aH|_\text{new,i}}{k}\right)^\frac{2}{1+3w}\left(\frac{k}{aH|_\text{new,i}}\right)^{-1} \nonumber \\
	=&\,\frac{H_\text{pre}}{2\pi}\left(\frac{aH|_\text{pre,f}}{aH|_\text{new,i}}\right)^{-\frac{3(1+w)}{1+3w}}
	\left(\frac{k}{aH|_\text{pre,f}}\right)^{-\mathrm{Re}\nu_\text{pre}-\frac{2}{1+3w}\mathrm{Re}\nu_\text{osc}}.
\end{align}
Finally, with use of $aH\propto a^{-\frac{1+3w}{2}}\propto\rho^\frac{1+3w}{6(1+w)}$, one can obtain
\begin{align}
	\delta\varphi|_3&\sim\frac{H_\text{pre}}{2\pi}\left(\frac{\rho_\text{pre}}{\rho_\text{new}}\right)^{-1/2}
	\left(\frac{k}{aH|_\text{pre,f}}\right)^{-\mathrm{Re}\nu_\text{pre}-\frac{2}{1+3w}\mathrm{Re}\nu_\text{osc}} \nonumber \\
	&\simeq\frac{H_\text{new}}{2\pi}\left(\frac{k}{aH|_\text{pre,f}}\right)^{-\mathrm{Re}\nu_\text{pre}-\frac{2}{1+3w}\mathrm{Re}\nu_\text{osc}}.
\end{align}
Therefore if $\varphi$ is sufficiently massive both during the pre-inflation and the oscillation phase as 
$\mathrm{Re}\nu_\text{pre}=\mathrm{Re}\nu_\text{osc}=0$, the amplitude of $\delta\phi$ at the second horizon exit is simply given by 
$H_\text{new}/2\pi$ and the curvature perturbation generating at this time can be estimated in the standard way
as we studied as a first example in the main text.
On the other hand, if the Hubble induced mass is as small as $\mathrm{Re}\nu_\text{pre}\,\,\text{or}\,\,\mathrm{Re}\nu_\text{osc}>0$,
the spectrum of $\delta\varphi|_3$ can be strongly red-tilted. Furthermore the $k$ dependence of the amplitude of the longer wavelength modes 
$k<aH|_\text{new,i}$ at $a_\text{new,i}$ is simply given by the damping factor during the pre-inflation phase $k^{\frac{3}{2}-\mathrm{Re}\nu_\text{pre}}$
since those modes are equally superhorizon during the oscillation phase.
Therefore, for large $c_\text{pot}$ and small $c$ (or equivalently small $\mathrm{Re}\nu_\text{pre}$ and large $\mathrm{Re}\nu_\text{osc}$),
a sharp peak of $\delta\varphi$ can be obtained on $k\sim aH|_\text{new,i}$ as a second example in the main text.

\section{Double Inflation in SUGRA Framework}
\label{sec:sugra}

Here we present a concrete supergravity (SUGRA) model 
as an origin of the potential given in the main text,
following Refs.~\cite{Kawasaki:2000yn,Kawasaki:2016pql}. We use the natural unit $\Mpl=1$ in this section for simplicity.

The model is based on discrete $R$ symmetry $Z_{2nR}$ which is broken down to a discrete $Z_{2R}$ during and after the second new inflation.
The $R$-invariant super- and K\"ahler potential are given by
\begin{align}
	W=&v^2\Phi-\frac{g}{n+1}\Phi^{n+1}+mSX+W_0, \\
	K=&\,|\Phi|^2+\frac{1}{2}\left(S+S^*\right)^2+|X|^2 \nonumber \\
	&+\frac{\kappa}{4}|\Phi|^4+\lambda|\Phi|^2|X|^2+\frac{\xi}{2}|\Phi|^2\left(S+S^*\right)^2.
\end{align}
Here $S$ and $\Phi$ include the inflatons $\phi$ and $\varphi$ for chaotic and new inflation respectively as their scalar components, and $X$ is called the \emph{stabilizer}.
Also we include a constant term $W_0$ in the superpotential.
Here $\Phi$ and $X$ are assumed to have $R$ charge 2, and $S$ and $X$ have additional $Z_2$ charge 1.
The K\"ahler potential is written by invariant terms under this symmetry, including relevant higher order terms.
In addition, we suppose a shift symmetry for $S$ so that the K\"ahler potential is invariant under the shift $\mathrm{Im}S\to\mathrm{Im}S+\alpha$
for an arbitrary $\alpha$.

For these super and K\"ahler potentials, the scalar potential is given by
\begin{align}
	V=\ee^K\left(K^{\bar{j}i}D_iWD_{\bar{j}}W^*-3|W|^2\right),
\end{align}
where bars denote complex conjugation, $K^{\bar{j}i}$ is the inverse of the K\"ahler metric 
$K_{i\bar{j}}=\partial^2K/\partial\phi^i\partial\bar{\phi}^{\bar{j}}$, and $D_i$ is a covariant derivative $D_iW=W_i+K_iW$. 
Then, defining the inflaton for chaotic and new inflation by $\varphi=\sqrt{2}\mathrm{Re}\Phi$ and $\phi=\sqrt{2}\mathrm{Im}S$,\footnote{$\Phi$
has $n$ potential minima and $\varphi=\mathrm{Re}\Phi$ is always the direction for one of them. Furthermore now we have a linear potential term
due to the constant superpotential $W_0$ which chose a $\varphi$'s positive direction.}
and assuming other degrees of freedom are stabilized at their origin due to the Hubble induced mass, 
the leading terms of the potential are given by
\begin{align}
	V\simeq&\, v^4-2\sqrt{2}W_0v^2\varphi-\frac{\kappa}{2}v^4\varphi^2-\frac{g}{2^{\frac{n}{2}-1}}v^2\varphi^n+\frac{g}{2^n}\varphi^{2n} \nonumber \\
	&+\frac{1}{2}m^2\phi^2\left(1+\frac{1-\lambda}{2}\varphi^2\right).
\end{align}
On the other hand, the kinetic terms are given by
\begin{align}
	\mathcal{L}_\mathrm{kin}=-K_{i\bar{j}}\partial_\mu\bar{\phi}^{\bar{j}}\partial^\mu\phi^i.
\end{align}
For relevant fields $\varphi$ and $\chi$, it reads
\begin{align}
	\mathcal{L}_\mathrm{kin}\supset-\frac{1}{2}\left(1+\frac{\kappa}{2}\varphi^2\right)(\partial\varphi)^2
	-\frac{1}{2}\left(1+\frac{\xi}{2}\varphi^2\right)(\partial\phi)^2.
\end{align}
Then the model which we considered in the main text has been derived.\footnote{In Eq.~(\ref{eq:kin_term}),
we omitted $\varphi^2(\partial\varphi)^2$ term since it is Planck- and slow-roll suppressed and almost irrelevant to the dynamics.}

%%%%%%%%%%%%%%%%%%%%%%%%%%%%%%%%%
%%%%%%%%%%% References %%%%%%%%%%%
%%%%%%%%%%%%%%%%%%%%%%%%%%%%%%%%%
\small
\bibliographystyle{apsrev4-1}
\bibliography{pbh_ligo}

\begin{thebibliography}{76}%
\makeatletter
\providecommand \@ifxundefined [1]{%
 \@ifx{#1\undefined}
}%
\providecommand \@ifnum [1]{%
 \ifnum #1\expandafter \@firstoftwo
 \else \expandafter \@secondoftwo
 \fi
}%
\providecommand \@ifx [1]{%
 \ifx #1\expandafter \@firstoftwo
 \else \expandafter \@secondoftwo
 \fi
}%
\providecommand \natexlab [1]{#1}%
\providecommand \enquote  [1]{``#1''}%
\providecommand \bibnamefont  [1]{#1}%
\providecommand \bibfnamefont [1]{#1}%
\providecommand \citenamefont [1]{#1}%
\providecommand \href@noop [0]{\@secondoftwo}%
\providecommand \href [0]{\begingroup \@sanitize@url \@href}%
\providecommand \@href[1]{\@@startlink{#1}\@@href}%
\providecommand \@@href[1]{\endgroup#1\@@endlink}%
\providecommand \@sanitize@url [0]{\catcode `\\12\catcode `\$12\catcode
  `\&12\catcode `\#12\catcode `\^12\catcode `\_12\catcode `\%12\relax}%
\providecommand \@@startlink[1]{}%
\providecommand \@@endlink[0]{}%
\providecommand \url  [0]{\begingroup\@sanitize@url \@url }%
\providecommand \@url [1]{\endgroup\@href {#1}{\urlprefix }}%
\providecommand \urlprefix  [0]{URL }%
\providecommand \Eprint [0]{\href }%
\providecommand \doibase [0]{http://dx.doi.org/}%
\providecommand \selectlanguage [0]{\@gobble}%
\providecommand \bibinfo  [0]{\@secondoftwo}%
\providecommand \bibfield  [0]{\@secondoftwo}%
\providecommand \translation [1]{[#1]}%
\providecommand \BibitemOpen [0]{}%
\providecommand \bibitemStop [0]{}%
\providecommand \bibitemNoStop [0]{.\EOS\space}%
\providecommand \EOS [0]{\spacefactor3000\relax}%
\providecommand \BibitemShut  [1]{\csname bibitem#1\endcsname}%
\let\auto@bib@innerbib\@empty
%</preamble>
\bibitem [{\citenamefont {Abbott}\ \emph
  {et~al.}(2016{\natexlab{a}})\citenamefont {Abbott} \emph
  {et~al.}}]{Abbott:2016blz}%
  \BibitemOpen
  \bibfield  {author} {\bibinfo {author} {\bibfnamefont {B.~P.}\ \bibnamefont
  {Abbott}} \emph {et~al.} (\bibinfo {collaboration} {Virgo, LIGO
  Scientific}),\ }\href {\doibase 10.1103/PhysRevLett.116.061102} {\bibfield
  {journal} {\bibinfo  {journal} {Phys. Rev. Lett.}\ }\textbf {\bibinfo
  {volume} {116}},\ \bibinfo {pages} {061102} (\bibinfo {year}
  {2016}{\natexlab{a}})},\ \Eprint {http://arxiv.org/abs/1602.03837}
  {arXiv:1602.03837 [gr-qc]} \BibitemShut {NoStop}%
%%CITATION = ARXIV:1602.03837;%%
\bibitem [{\citenamefont {Abbott}\ \emph
  {et~al.}(2016{\natexlab{b}})\citenamefont {Abbott} \emph
  {et~al.}}]{Abbott:2016nmj}%
  \BibitemOpen
  \bibfield  {author} {\bibinfo {author} {\bibfnamefont {B.~P.}\ \bibnamefont
  {Abbott}} \emph {et~al.} (\bibinfo {collaboration} {Virgo, LIGO
  Scientific}),\ }\href {\doibase 10.1103/PhysRevLett.116.241103} {\bibfield
  {journal} {\bibinfo  {journal} {Phys. Rev. Lett.}\ }\textbf {\bibinfo
  {volume} {116}},\ \bibinfo {pages} {241103} (\bibinfo {year}
  {2016}{\natexlab{b}})},\ \Eprint {http://arxiv.org/abs/1606.04855}
  {arXiv:1606.04855 [gr-qc]} \BibitemShut {NoStop}%
%%CITATION = ARXIV:1606.04855;%%
\bibitem [{\citenamefont {Abbott}\ \emph
  {et~al.}(2016{\natexlab{c}})\citenamefont {Abbott} \emph
  {et~al.}}]{TheLIGOScientific:2016pea}%
  \BibitemOpen
  \bibfield  {author} {\bibinfo {author} {\bibfnamefont {B.~P.}\ \bibnamefont
  {Abbott}} \emph {et~al.} (\bibinfo {collaboration} {Virgo, LIGO
  Scientific}),\ }\href {\doibase 10.1103/PhysRevX.6.041015} {\bibfield
  {journal} {\bibinfo  {journal} {Phys. Rev.}\ }\textbf {\bibinfo {volume}
  {X6}},\ \bibinfo {pages} {041015} (\bibinfo {year} {2016}{\natexlab{c}})},\
  \Eprint {http://arxiv.org/abs/1606.04856} {arXiv:1606.04856 [gr-qc]}
  \BibitemShut {NoStop}%
%%CITATION = ARXIV:1606.04856;%%
\bibitem [{\citenamefont {Bird}\ \emph {et~al.}(2016)\citenamefont {Bird},
  \citenamefont {Cholis}, \citenamefont {Mu{\~n}oz}, \citenamefont
  {Ali-Ha{\"i}moud}, \citenamefont {Kamionkowski}, \citenamefont {Kovetz},
  \citenamefont {Raccanelli},\ and\ \citenamefont {Riess}}]{Bird:2016dcv}%
  \BibitemOpen
  \bibfield  {author} {\bibinfo {author} {\bibfnamefont {S.}~\bibnamefont
  {Bird}}, \bibinfo {author} {\bibfnamefont {I.}~\bibnamefont {Cholis}},
  \bibinfo {author} {\bibfnamefont {J.~B.}\ \bibnamefont {Mu{\~n}oz}}, \bibinfo
  {author} {\bibfnamefont {Y.}~\bibnamefont {Ali-Ha{\"i}moud}}, \bibinfo
  {author} {\bibfnamefont {M.}~\bibnamefont {Kamionkowski}}, \bibinfo {author}
  {\bibfnamefont {E.~D.}\ \bibnamefont {Kovetz}}, \bibinfo {author}
  {\bibfnamefont {A.}~\bibnamefont {Raccanelli}}, \ and\ \bibinfo {author}
  {\bibfnamefont {A.~G.}\ \bibnamefont {Riess}},\ }\href {\doibase
  10.1103/PhysRevLett.116.201301} {\bibfield  {journal} {\bibinfo  {journal}
  {Phys. Rev. Lett.}\ }\textbf {\bibinfo {volume} {116}},\ \bibinfo {pages}
  {201301} (\bibinfo {year} {2016})},\ \Eprint
  {http://arxiv.org/abs/1603.00464} {arXiv:1603.00464 [astro-ph.CO]}
  \BibitemShut {NoStop}%
%%CITATION = ARXIV:1603.00464;%%
\bibitem [{\citenamefont {Clesse}\ and\ \citenamefont
  {Garc{\'i}a-Bellido}(2017)}]{Clesse:2016vqa}%
  \BibitemOpen
  \bibfield  {author} {\bibinfo {author} {\bibfnamefont {S.}~\bibnamefont
  {Clesse}}\ and\ \bibinfo {author} {\bibfnamefont {J.}~\bibnamefont
  {Garc{\'i}a-Bellido}},\ }\href {\doibase 10.1016/j.dark.2016.10.002}
  {\bibfield  {journal} {\bibinfo  {journal} {Phys. Dark Univ.}\ }\textbf
  {\bibinfo {volume} {15}},\ \bibinfo {pages} {142} (\bibinfo {year} {2017})},\
  \Eprint {http://arxiv.org/abs/1603.05234} {arXiv:1603.05234 [astro-ph.CO]}
  \BibitemShut {NoStop}%
%%CITATION = ARXIV:1603.05234;%%
\bibitem [{\citenamefont {Sasaki}\ \emph {et~al.}(2016)\citenamefont {Sasaki},
  \citenamefont {Suyama}, \citenamefont {Tanaka},\ and\ \citenamefont
  {Yokoyama}}]{Sasaki:2016jop}%
  \BibitemOpen
  \bibfield  {author} {\bibinfo {author} {\bibfnamefont {M.}~\bibnamefont
  {Sasaki}}, \bibinfo {author} {\bibfnamefont {T.}~\bibnamefont {Suyama}},
  \bibinfo {author} {\bibfnamefont {T.}~\bibnamefont {Tanaka}}, \ and\ \bibinfo
  {author} {\bibfnamefont {S.}~\bibnamefont {Yokoyama}},\ }\href {\doibase
  10.1103/PhysRevLett.117.061101} {\bibfield  {journal} {\bibinfo  {journal}
  {Phys. Rev. Lett.}\ }\textbf {\bibinfo {volume} {117}},\ \bibinfo {pages}
  {061101} (\bibinfo {year} {2016})},\ \Eprint
  {http://arxiv.org/abs/1603.08338} {arXiv:1603.08338 [astro-ph.CO]}
  \BibitemShut {NoStop}%
%%CITATION = ARXIV:1603.08338;%%
\bibitem [{\citenamefont {Eroshenko}(2016)}]{Eroshenko:2016hmn}%
  \BibitemOpen
  \bibfield  {author} {\bibinfo {author} {\bibfnamefont {{\relax Yu}.~N.}\
  \bibnamefont {Eroshenko}},\ }\href@noop {} {\  (\bibinfo {year} {2016})},\
  \Eprint {http://arxiv.org/abs/1604.04932} {arXiv:1604.04932 [astro-ph.CO]}
  \BibitemShut {NoStop}%
%%CITATION = ARXIV:1604.04932;%%
\bibitem [{\citenamefont {Carr}\ \emph {et~al.}(2016)\citenamefont {Carr},
  \citenamefont {Kuhnel},\ and\ \citenamefont {Sandstad}}]{Carr:2016drx}%
  \BibitemOpen
  \bibfield  {author} {\bibinfo {author} {\bibfnamefont {B.}~\bibnamefont
  {Carr}}, \bibinfo {author} {\bibfnamefont {F.}~\bibnamefont {Kuhnel}}, \ and\
  \bibinfo {author} {\bibfnamefont {M.}~\bibnamefont {Sandstad}},\ }\href
  {\doibase 10.1103/PhysRevD.94.083504} {\bibfield  {journal} {\bibinfo
  {journal} {Phys. Rev.}\ }\textbf {\bibinfo {volume} {D94}},\ \bibinfo {pages}
  {083504} (\bibinfo {year} {2016})},\ \Eprint
  {http://arxiv.org/abs/1607.06077} {arXiv:1607.06077 [astro-ph.CO]}
  \BibitemShut {NoStop}%
%%CITATION = ARXIV:1607.06077;%%
\bibitem [{\citenamefont {Hawking}(1971)}]{Hawking:1971ei}%
  \BibitemOpen
  \bibfield  {author} {\bibinfo {author} {\bibfnamefont {S.}~\bibnamefont
  {Hawking}},\ }\href@noop {} {\bibfield  {journal} {\bibinfo  {journal} {Mon.
  Not. Roy. Astron. Soc.}\ }\textbf {\bibinfo {volume} {152}},\ \bibinfo
  {pages} {75} (\bibinfo {year} {1971})}\BibitemShut {NoStop}%
%%CITATION = MNRAA,152,75;%%
\bibitem [{\citenamefont {Carr}\ and\ \citenamefont
  {Hawking}(1974)}]{Carr:1974nx}%
  \BibitemOpen
  \bibfield  {author} {\bibinfo {author} {\bibfnamefont {B.~J.}\ \bibnamefont
  {Carr}}\ and\ \bibinfo {author} {\bibfnamefont {S.~W.}\ \bibnamefont
  {Hawking}},\ }\href@noop {} {\bibfield  {journal} {\bibinfo  {journal} {Mon.
  Not. Roy. Astron. Soc.}\ }\textbf {\bibinfo {volume} {168}},\ \bibinfo
  {pages} {399} (\bibinfo {year} {1974})}\BibitemShut {NoStop}%
%%CITATION = MNRAA,168,399;%%
\bibitem [{\citenamefont {Carr}(1975)}]{Carr:1975qj}%
  \BibitemOpen
  \bibfield  {author} {\bibinfo {author} {\bibfnamefont {B.~J.}\ \bibnamefont
  {Carr}},\ }\href {\doibase 10.1086/153853} {\bibfield  {journal} {\bibinfo
  {journal} {Astrophys. J.}\ }\textbf {\bibinfo {volume} {201}},\ \bibinfo
  {pages} {1} (\bibinfo {year} {1975})}\BibitemShut {NoStop}%
%%CITATION = ASJOA,201,1;%%
\bibitem [{\citenamefont {Garcia-Bellido}\ \emph {et~al.}(1996)\citenamefont
  {Garcia-Bellido}, \citenamefont {Linde},\ and\ \citenamefont
  {Wands}}]{GarciaBellido:1996qt}%
  \BibitemOpen
  \bibfield  {author} {\bibinfo {author} {\bibfnamefont {J.}~\bibnamefont
  {Garcia-Bellido}}, \bibinfo {author} {\bibfnamefont {A.~D.}\ \bibnamefont
  {Linde}}, \ and\ \bibinfo {author} {\bibfnamefont {D.}~\bibnamefont
  {Wands}},\ }\href {\doibase 10.1103/PhysRevD.54.6040} {\bibfield  {journal}
  {\bibinfo  {journal} {Phys. Rev.}\ }\textbf {\bibinfo {volume} {D54}},\
  \bibinfo {pages} {6040} (\bibinfo {year} {1996})},\ \Eprint
  {http://arxiv.org/abs/astro-ph/9605094} {arXiv:astro-ph/9605094 [astro-ph]}
  \BibitemShut {NoStop}%
%%CITATION = ASTRO-PH/9605094;%%
\bibitem [{\citenamefont {Kawasaki}\ \emph {et~al.}(1998)\citenamefont
  {Kawasaki}, \citenamefont {Sugiyama},\ and\ \citenamefont
  {Yanagida}}]{Kawasaki:1997ju}%
  \BibitemOpen
  \bibfield  {author} {\bibinfo {author} {\bibfnamefont {M.}~\bibnamefont
  {Kawasaki}}, \bibinfo {author} {\bibfnamefont {N.}~\bibnamefont {Sugiyama}},
  \ and\ \bibinfo {author} {\bibfnamefont {T.}~\bibnamefont {Yanagida}},\
  }\href {\doibase 10.1103/PhysRevD.57.6050} {\bibfield  {journal} {\bibinfo
  {journal} {Phys. Rev.}\ }\textbf {\bibinfo {volume} {D57}},\ \bibinfo {pages}
  {6050} (\bibinfo {year} {1998})},\ \Eprint
  {http://arxiv.org/abs/hep-ph/9710259} {arXiv:hep-ph/9710259 [hep-ph]}
  \BibitemShut {NoStop}%
%%CITATION = HEP-PH/9710259;%%
\bibitem [{\citenamefont {Yokoyama}(1998)}]{Yokoyama:1998pt}%
  \BibitemOpen
  \bibfield  {author} {\bibinfo {author} {\bibfnamefont {J.}~\bibnamefont
  {Yokoyama}},\ }\href {\doibase 10.1103/PhysRevD.58.083510} {\bibfield
  {journal} {\bibinfo  {journal} {Phys. Rev.}\ }\textbf {\bibinfo {volume}
  {D58}},\ \bibinfo {pages} {083510} (\bibinfo {year} {1998})},\ \Eprint
  {http://arxiv.org/abs/astro-ph/9802357} {arXiv:astro-ph/9802357 [astro-ph]}
  \BibitemShut {NoStop}%
%%CITATION = ASTRO-PH/9802357;%%
\bibitem [{\citenamefont {Chluba}\ \emph {et~al.}(2012)\citenamefont {Chluba},
  \citenamefont {Erickcek},\ and\ \citenamefont {Ben-Dayan}}]{Chluba:2012we}%
  \BibitemOpen
  \bibfield  {author} {\bibinfo {author} {\bibfnamefont {J.}~\bibnamefont
  {Chluba}}, \bibinfo {author} {\bibfnamefont {A.~L.}\ \bibnamefont
  {Erickcek}}, \ and\ \bibinfo {author} {\bibfnamefont {I.}~\bibnamefont
  {Ben-Dayan}},\ }\href {\doibase 10.1088/0004-637X/758/2/76} {\bibfield
  {journal} {\bibinfo  {journal} {Astrophys. J.}\ }\textbf {\bibinfo {volume}
  {758}},\ \bibinfo {pages} {76} (\bibinfo {year} {2012})},\ \Eprint
  {http://arxiv.org/abs/1203.2681} {arXiv:1203.2681 [astro-ph.CO]} \BibitemShut
  {NoStop}%
%%CITATION = ARXIV:1203.2681;%%
\bibitem [{\citenamefont {Kohri}\ \emph {et~al.}(2014)\citenamefont {Kohri},
  \citenamefont {Nakama},\ and\ \citenamefont {Suyama}}]{Kohri:2014lza}%
  \BibitemOpen
  \bibfield  {author} {\bibinfo {author} {\bibfnamefont {K.}~\bibnamefont
  {Kohri}}, \bibinfo {author} {\bibfnamefont {T.}~\bibnamefont {Nakama}}, \
  and\ \bibinfo {author} {\bibfnamefont {T.}~\bibnamefont {Suyama}},\ }\href
  {\doibase 10.1103/PhysRevD.90.083514} {\bibfield  {journal} {\bibinfo
  {journal} {Phys. Rev.}\ }\textbf {\bibinfo {volume} {D90}},\ \bibinfo {pages}
  {083514} (\bibinfo {year} {2014})},\ \Eprint {http://arxiv.org/abs/1405.5999}
  {arXiv:1405.5999 [astro-ph.CO]} \BibitemShut {NoStop}%
%%CITATION = ARXIV:1405.5999;%%
\bibitem [{\citenamefont {Saito}\ and\ \citenamefont
  {Yokoyama}(2009)}]{Saito:2008jc}%
  \BibitemOpen
  \bibfield  {author} {\bibinfo {author} {\bibfnamefont {R.}~\bibnamefont
  {Saito}}\ and\ \bibinfo {author} {\bibfnamefont {J.}~\bibnamefont
  {Yokoyama}},\ }\href {\doibase 10.1103/PhysRevLett.102.161101,
  10.1103/PhysRevLett.107.069901} {\bibfield  {journal} {\bibinfo  {journal}
  {Phys. Rev. Lett.}\ }\textbf {\bibinfo {volume} {102}},\ \bibinfo {pages}
  {161101} (\bibinfo {year} {2009})},\ \bibinfo {note} {[Erratum: Phys. Rev.
  Lett.107,069901(2011)]},\ \Eprint {http://arxiv.org/abs/0812.4339}
  {arXiv:0812.4339 [astro-ph]} \BibitemShut {NoStop}%
%%CITATION = ARXIV:0812.4339;%%
\bibitem [{\citenamefont {Saito}\ and\ \citenamefont
  {Yokoyama}(2010)}]{Saito:2009jt}%
  \BibitemOpen
  \bibfield  {author} {\bibinfo {author} {\bibfnamefont {R.}~\bibnamefont
  {Saito}}\ and\ \bibinfo {author} {\bibfnamefont {J.}~\bibnamefont
  {Yokoyama}},\ }\href {\doibase 10.1143/PTP.126.351, 10.1143/PTP.123.867}
  {\bibfield  {journal} {\bibinfo  {journal} {Prog. Theor. Phys.}\ }\textbf
  {\bibinfo {volume} {123}},\ \bibinfo {pages} {867} (\bibinfo {year}
  {2010})},\ \bibinfo {note} {[Erratum: Prog. Theor. Phys.126,351(2011)]},\
  \Eprint {http://arxiv.org/abs/0912.5317} {arXiv:0912.5317 [astro-ph.CO]}
  \BibitemShut {NoStop}%
%%CITATION = ARXIV:0912.5317;%%
\bibitem [{\citenamefont {Bugaev}\ and\ \citenamefont
  {Klimai}(2010)}]{Bugaev:2009zh}%
  \BibitemOpen
  \bibfield  {author} {\bibinfo {author} {\bibfnamefont {E.}~\bibnamefont
  {Bugaev}}\ and\ \bibinfo {author} {\bibfnamefont {P.}~\bibnamefont
  {Klimai}},\ }\href {\doibase 10.1103/PhysRevD.81.023517} {\bibfield
  {journal} {\bibinfo  {journal} {Phys. Rev.}\ }\textbf {\bibinfo {volume}
  {D81}},\ \bibinfo {pages} {023517} (\bibinfo {year} {2010})},\ \Eprint
  {http://arxiv.org/abs/0908.0664} {arXiv:0908.0664 [astro-ph.CO]} \BibitemShut
  {NoStop}%
%%CITATION = ARXIV:0908.0664;%%
\bibitem [{\citenamefont {Bugaev}\ and\ \citenamefont
  {Klimai}(2011)}]{Bugaev:2010bb}%
  \BibitemOpen
  \bibfield  {author} {\bibinfo {author} {\bibfnamefont {E.}~\bibnamefont
  {Bugaev}}\ and\ \bibinfo {author} {\bibfnamefont {P.}~\bibnamefont
  {Klimai}},\ }\href {\doibase 10.1103/PhysRevD.83.083521} {\bibfield
  {journal} {\bibinfo  {journal} {Phys. Rev.}\ }\textbf {\bibinfo {volume}
  {D83}},\ \bibinfo {pages} {083521} (\bibinfo {year} {2011})},\ \Eprint
  {http://arxiv.org/abs/1012.4697} {arXiv:1012.4697 [astro-ph.CO]} \BibitemShut
  {NoStop}%
%%CITATION = ARXIV:1012.4697;%%
\bibitem [{\citenamefont {Ananda}\ \emph {et~al.}(2007)\citenamefont {Ananda},
  \citenamefont {Clarkson},\ and\ \citenamefont {Wands}}]{Ananda:2006af}%
  \BibitemOpen
  \bibfield  {author} {\bibinfo {author} {\bibfnamefont {K.~N.}\ \bibnamefont
  {Ananda}}, \bibinfo {author} {\bibfnamefont {C.}~\bibnamefont {Clarkson}}, \
  and\ \bibinfo {author} {\bibfnamefont {D.}~\bibnamefont {Wands}},\ }\href
  {\doibase 10.1103/PhysRevD.75.123518} {\bibfield  {journal} {\bibinfo
  {journal} {Phys. Rev.}\ }\textbf {\bibinfo {volume} {D75}},\ \bibinfo {pages}
  {123518} (\bibinfo {year} {2007})},\ \Eprint
  {http://arxiv.org/abs/gr-qc/0612013} {arXiv:gr-qc/0612013 [gr-qc]}
  \BibitemShut {NoStop}%
%%CITATION = GR-QC/0612013;%%
\bibitem [{\citenamefont {Baumann}\ \emph {et~al.}(2007)\citenamefont
  {Baumann}, \citenamefont {Steinhardt}, \citenamefont {Takahashi},\ and\
  \citenamefont {Ichiki}}]{Baumann:2007zm}%
  \BibitemOpen
  \bibfield  {author} {\bibinfo {author} {\bibfnamefont {D.}~\bibnamefont
  {Baumann}}, \bibinfo {author} {\bibfnamefont {P.~J.}\ \bibnamefont
  {Steinhardt}}, \bibinfo {author} {\bibfnamefont {K.}~\bibnamefont
  {Takahashi}}, \ and\ \bibinfo {author} {\bibfnamefont {K.}~\bibnamefont
  {Ichiki}},\ }\href {\doibase 10.1103/PhysRevD.76.084019} {\bibfield
  {journal} {\bibinfo  {journal} {Phys. Rev.}\ }\textbf {\bibinfo {volume}
  {D76}},\ \bibinfo {pages} {084019} (\bibinfo {year} {2007})},\ \Eprint
  {http://arxiv.org/abs/hep-th/0703290} {arXiv:hep-th/0703290 [hep-th]}
  \BibitemShut {NoStop}%
%%CITATION = HEP-TH/0703290;%%
\bibitem [{\citenamefont {Arzoumanian}\ \emph {et~al.}(2016)\citenamefont
  {Arzoumanian} \emph {et~al.}}]{Arzoumanian:2015liz}%
  \BibitemOpen
  \bibfield  {author} {\bibinfo {author} {\bibfnamefont {Z.}~\bibnamefont
  {Arzoumanian}} \emph {et~al.} (\bibinfo {collaboration} {NANOGrav}),\ }\href
  {\doibase 10.3847/0004-637X/821/1/13} {\bibfield  {journal} {\bibinfo
  {journal} {Astrophys. J.}\ }\textbf {\bibinfo {volume} {821}},\ \bibinfo
  {pages} {13} (\bibinfo {year} {2016})},\ \Eprint
  {http://arxiv.org/abs/1508.03024} {arXiv:1508.03024 [astro-ph.GA]}
  \BibitemShut {NoStop}%
%%CITATION = ARXIV:1508.03024;%%
\bibitem [{\citenamefont {Lentati}\ \emph {et~al.}(2015)\citenamefont {Lentati}
  \emph {et~al.}}]{Lentati:2015qwp}%
  \BibitemOpen
  \bibfield  {author} {\bibinfo {author} {\bibfnamefont {L.}~\bibnamefont
  {Lentati}} \emph {et~al.},\ }\href {\doibase 10.1093/mnras/stv1538}
  {\bibfield  {journal} {\bibinfo  {journal} {Mon. Not. Roy. Astron. Soc.}\
  }\textbf {\bibinfo {volume} {453}},\ \bibinfo {pages} {2576} (\bibinfo {year}
  {2015})},\ \Eprint {http://arxiv.org/abs/1504.03692} {arXiv:1504.03692
  [astro-ph.CO]} \BibitemShut {NoStop}%
%%CITATION = ARXIV:1504.03692;%%
\bibitem [{\citenamefont {Shannon}\ \emph {et~al.}(2015)\citenamefont {Shannon}
  \emph {et~al.}}]{Shannon:2015ect}%
  \BibitemOpen
  \bibfield  {author} {\bibinfo {author} {\bibfnamefont {R.~M.}\ \bibnamefont
  {Shannon}} \emph {et~al.},\ }\href {\doibase 10.1126/science.aab1910}
  {\bibfield  {journal} {\bibinfo  {journal} {Science}\ }\textbf {\bibinfo
  {volume} {349}},\ \bibinfo {pages} {1522} (\bibinfo {year} {2015})},\ \Eprint
  {http://arxiv.org/abs/1509.07320} {arXiv:1509.07320 [astro-ph.CO]}
  \BibitemShut {NoStop}%
%%CITATION = ARXIV:1509.07320;%%
\bibitem [{\citenamefont {Schutz}\ and\ \citenamefont
  {Liu}(2017)}]{Schutz:2016khr}%
  \BibitemOpen
  \bibfield  {author} {\bibinfo {author} {\bibfnamefont {K.}~\bibnamefont
  {Schutz}}\ and\ \bibinfo {author} {\bibfnamefont {A.}~\bibnamefont {Liu}},\
  }\href {\doibase 10.1103/PhysRevD.95.023002} {\bibfield  {journal} {\bibinfo
  {journal} {Phys. Rev.}\ }\textbf {\bibinfo {volume} {D95}},\ \bibinfo {pages}
  {023002} (\bibinfo {year} {2017})},\ \Eprint
  {http://arxiv.org/abs/1610.04234} {arXiv:1610.04234 [astro-ph.CO]}
  \BibitemShut {NoStop}%
%%CITATION = ARXIV:1610.04234;%%
\bibitem [{\citenamefont {Clesse}\ and\ \citenamefont
  {Garc{\'i}a-Bellido}(2016)}]{Clesse:2016ajp}%
  \BibitemOpen
  \bibfield  {author} {\bibinfo {author} {\bibfnamefont {S.}~\bibnamefont
  {Clesse}}\ and\ \bibinfo {author} {\bibfnamefont {J.}~\bibnamefont
  {Garc{\'i}a-Bellido}},\ }\href@noop {} {\  (\bibinfo {year} {2016})},\
  \Eprint {http://arxiv.org/abs/1610.08479} {arXiv:1610.08479 [astro-ph.CO]}
  \BibitemShut {NoStop}%
%%CITATION = ARXIV:1610.08479;%%
\bibitem [{\citenamefont {Kawasaki}\ and\ \citenamefont
  {Yanagida}(1999)}]{Kawasaki:1998vx}%
  \BibitemOpen
  \bibfield  {author} {\bibinfo {author} {\bibfnamefont {M.}~\bibnamefont
  {Kawasaki}}\ and\ \bibinfo {author} {\bibfnamefont {T.}~\bibnamefont
  {Yanagida}},\ }\href {\doibase 10.1103/PhysRevD.59.043512} {\bibfield
  {journal} {\bibinfo  {journal} {Phys. Rev.}\ }\textbf {\bibinfo {volume}
  {D59}},\ \bibinfo {pages} {043512} (\bibinfo {year} {1999})},\ \Eprint
  {http://arxiv.org/abs/hep-ph/9807544} {arXiv:hep-ph/9807544 [hep-ph]}
  \BibitemShut {NoStop}%
%%CITATION = HEP-PH/9807544;%%
\bibitem [{\citenamefont {Frampton}\ \emph {et~al.}(2010)\citenamefont
  {Frampton}, \citenamefont {Kawasaki}, \citenamefont {Takahashi},\ and\
  \citenamefont {Yanagida}}]{Frampton:2010sw}%
  \BibitemOpen
  \bibfield  {author} {\bibinfo {author} {\bibfnamefont {P.~H.}\ \bibnamefont
  {Frampton}}, \bibinfo {author} {\bibfnamefont {M.}~\bibnamefont {Kawasaki}},
  \bibinfo {author} {\bibfnamefont {F.}~\bibnamefont {Takahashi}}, \ and\
  \bibinfo {author} {\bibfnamefont {T.~T.}\ \bibnamefont {Yanagida}},\ }\href
  {\doibase 10.1088/1475-7516/2010/04/023} {\bibfield  {journal} {\bibinfo
  {journal} {JCAP}\ }\textbf {\bibinfo {volume} {1004}},\ \bibinfo {pages}
  {023} (\bibinfo {year} {2010})},\ \Eprint {http://arxiv.org/abs/1001.2308}
  {arXiv:1001.2308 [hep-ph]} \BibitemShut {NoStop}%
%%CITATION = ARXIV:1001.2308;%%
\bibitem [{\citenamefont {Kawasaki}\ \emph {et~al.}(2012)\citenamefont
  {Kawasaki}, \citenamefont {Kusenko},\ and\ \citenamefont
  {Yanagida}}]{Kawasaki:2012kn}%
  \BibitemOpen
  \bibfield  {author} {\bibinfo {author} {\bibfnamefont {M.}~\bibnamefont
  {Kawasaki}}, \bibinfo {author} {\bibfnamefont {A.}~\bibnamefont {Kusenko}}, \
  and\ \bibinfo {author} {\bibfnamefont {T.~T.}\ \bibnamefont {Yanagida}},\
  }\href {\doibase 10.1016/j.physletb.2012.03.056} {\bibfield  {journal}
  {\bibinfo  {journal} {Phys. Lett.}\ }\textbf {\bibinfo {volume} {B711}},\
  \bibinfo {pages} {1} (\bibinfo {year} {2012})},\ \Eprint
  {http://arxiv.org/abs/1202.3848} {arXiv:1202.3848 [astro-ph.CO]} \BibitemShut
  {NoStop}%
%%CITATION = ARXIV:1202.3848;%%
\bibitem [{\citenamefont {Kawasaki}\ \emph
  {et~al.}(2016{\natexlab{a}})\citenamefont {Kawasaki}, \citenamefont
  {Mukaida},\ and\ \citenamefont {Yanagida}}]{Kawasaki:2016ijp}%
  \BibitemOpen
  \bibfield  {author} {\bibinfo {author} {\bibfnamefont {M.}~\bibnamefont
  {Kawasaki}}, \bibinfo {author} {\bibfnamefont {K.}~\bibnamefont {Mukaida}}, \
  and\ \bibinfo {author} {\bibfnamefont {T.~T.}\ \bibnamefont {Yanagida}},\
  }\href {\doibase 10.1103/PhysRevD.94.063509} {\bibfield  {journal} {\bibinfo
  {journal} {Phys. Rev.}\ }\textbf {\bibinfo {volume} {D94}},\ \bibinfo {pages}
  {063509} (\bibinfo {year} {2016}{\natexlab{a}})},\ \Eprint
  {http://arxiv.org/abs/1605.04974} {arXiv:1605.04974 [hep-ph]} \BibitemShut
  {NoStop}%
%%CITATION = ARXIV:1605.04974;%%
\bibitem [{\citenamefont {Kawasaki}\ \emph
  {et~al.}(2016{\natexlab{b}})\citenamefont {Kawasaki}, \citenamefont
  {Kusenko}, \citenamefont {Tada},\ and\ \citenamefont
  {Yanagida}}]{Kawasaki:2016pql}%
  \BibitemOpen
  \bibfield  {author} {\bibinfo {author} {\bibfnamefont {M.}~\bibnamefont
  {Kawasaki}}, \bibinfo {author} {\bibfnamefont {A.}~\bibnamefont {Kusenko}},
  \bibinfo {author} {\bibfnamefont {Y.}~\bibnamefont {Tada}}, \ and\ \bibinfo
  {author} {\bibfnamefont {T.~T.}\ \bibnamefont {Yanagida}},\ }\href {\doibase
  10.1103/PhysRevD.94.083523} {\bibfield  {journal} {\bibinfo  {journal} {Phys.
  Rev.}\ }\textbf {\bibinfo {volume} {D94}},\ \bibinfo {pages} {083523}
  (\bibinfo {year} {2016}{\natexlab{b}})},\ \Eprint
  {http://arxiv.org/abs/1606.07631} {arXiv:1606.07631 [astro-ph.CO]}
  \BibitemShut {NoStop}%
%%CITATION = ARXIV:1606.07631;%%
\bibitem [{\citenamefont {Green}\ and\ \citenamefont
  {Liddle}(1997)}]{Green:1997sz}%
  \BibitemOpen
  \bibfield  {author} {\bibinfo {author} {\bibfnamefont {A.~M.}\ \bibnamefont
  {Green}}\ and\ \bibinfo {author} {\bibfnamefont {A.~R.}\ \bibnamefont
  {Liddle}},\ }\href {\doibase 10.1103/PhysRevD.56.6166} {\bibfield  {journal}
  {\bibinfo  {journal} {Phys. Rev.}\ }\textbf {\bibinfo {volume} {D56}},\
  \bibinfo {pages} {6166} (\bibinfo {year} {1997})},\ \Eprint
  {http://arxiv.org/abs/astro-ph/9704251} {arXiv:astro-ph/9704251 [astro-ph]}
  \BibitemShut {NoStop}%
%%CITATION = ASTRO-PH/9704251;%%
\bibitem [{\citenamefont {Choptuik}(1993)}]{Choptuik:1992jv}%
  \BibitemOpen
  \bibfield  {author} {\bibinfo {author} {\bibfnamefont {M.~W.}\ \bibnamefont
  {Choptuik}},\ }\href {\doibase 10.1103/PhysRevLett.70.9} {\bibfield
  {journal} {\bibinfo  {journal} {Phys. Rev. Lett.}\ }\textbf {\bibinfo
  {volume} {70}},\ \bibinfo {pages} {9} (\bibinfo {year} {1993})}\BibitemShut
  {NoStop}%
%%CITATION = PRLTA,70,9;%%
\bibitem [{\citenamefont {Niemeyer}\ and\ \citenamefont
  {Jedamzik}(1998)}]{Niemeyer:1997mt}%
  \BibitemOpen
  \bibfield  {author} {\bibinfo {author} {\bibfnamefont {J.~C.}\ \bibnamefont
  {Niemeyer}}\ and\ \bibinfo {author} {\bibfnamefont {K.}~\bibnamefont
  {Jedamzik}},\ }\href {\doibase 10.1103/PhysRevLett.80.5481} {\bibfield
  {journal} {\bibinfo  {journal} {Phys. Rev. Lett.}\ }\textbf {\bibinfo
  {volume} {80}},\ \bibinfo {pages} {5481} (\bibinfo {year} {1998})},\ \Eprint
  {http://arxiv.org/abs/astro-ph/9709072} {arXiv:astro-ph/9709072 [astro-ph]}
  \BibitemShut {NoStop}%
%%CITATION = ASTRO-PH/9709072;%%
\bibitem [{\citenamefont {Niemeyer}(1998)}]{Niemeyer:1998ac}%
  \BibitemOpen
  \bibfield  {author} {\bibinfo {author} {\bibfnamefont {J.~C.}\ \bibnamefont
  {Niemeyer}},\ }in\ \href@noop {} {\emph {\bibinfo {booktitle} {{Sources and
  detection of dark matter in the universe. Proceedings, 3rd International
  Symposium, and Workshop on Primordial Black Holes and Hawking Radiation,
  Marina del Rey, USA, February 17-20, 1998}}}}\ (\bibinfo {year} {1998})\
  \Eprint {http://arxiv.org/abs/astro-ph/9806043} {arXiv:astro-ph/9806043
  [astro-ph]} \BibitemShut {NoStop}%
%%CITATION = ASTRO-PH/9806043;%%
\bibitem [{\citenamefont {Niemeyer}\ and\ \citenamefont
  {Jedamzik}(1999)}]{Niemeyer:1999ak}%
  \BibitemOpen
  \bibfield  {author} {\bibinfo {author} {\bibfnamefont {J.~C.}\ \bibnamefont
  {Niemeyer}}\ and\ \bibinfo {author} {\bibfnamefont {K.}~\bibnamefont
  {Jedamzik}},\ }\href {\doibase 10.1103/PhysRevD.59.124013} {\bibfield
  {journal} {\bibinfo  {journal} {Phys. Rev.}\ }\textbf {\bibinfo {volume}
  {D59}},\ \bibinfo {pages} {124013} (\bibinfo {year} {1999})},\ \Eprint
  {http://arxiv.org/abs/astro-ph/9901292} {arXiv:astro-ph/9901292 [astro-ph]}
  \BibitemShut {NoStop}%
%%CITATION = ASTRO-PH/9901292;%%
\bibitem [{\citenamefont {Musco}\ \emph {et~al.}(2005)\citenamefont {Musco},
  \citenamefont {Miller},\ and\ \citenamefont {Rezzolla}}]{Musco:2004ak}%
  \BibitemOpen
  \bibfield  {author} {\bibinfo {author} {\bibfnamefont {I.}~\bibnamefont
  {Musco}}, \bibinfo {author} {\bibfnamefont {J.~C.}\ \bibnamefont {Miller}}, \
  and\ \bibinfo {author} {\bibfnamefont {L.}~\bibnamefont {Rezzolla}},\ }\href
  {\doibase 10.1088/0264-9381/22/7/013} {\bibfield  {journal} {\bibinfo
  {journal} {Class. Quant. Grav.}\ }\textbf {\bibinfo {volume} {22}},\ \bibinfo
  {pages} {1405} (\bibinfo {year} {2005})},\ \Eprint
  {http://arxiv.org/abs/gr-qc/0412063} {arXiv:gr-qc/0412063 [gr-qc]}
  \BibitemShut {NoStop}%
%%CITATION = GR-QC/0412063;%%
\bibitem [{\citenamefont {Musco}\ \emph {et~al.}(2009)\citenamefont {Musco},
  \citenamefont {Miller},\ and\ \citenamefont {Polnarev}}]{Musco:2008hv}%
  \BibitemOpen
  \bibfield  {author} {\bibinfo {author} {\bibfnamefont {I.}~\bibnamefont
  {Musco}}, \bibinfo {author} {\bibfnamefont {J.~C.}\ \bibnamefont {Miller}}, \
  and\ \bibinfo {author} {\bibfnamefont {A.~G.}\ \bibnamefont {Polnarev}},\
  }\href {\doibase 10.1088/0264-9381/26/23/235001} {\bibfield  {journal}
  {\bibinfo  {journal} {Class. Quant. Grav.}\ }\textbf {\bibinfo {volume}
  {26}},\ \bibinfo {pages} {235001} (\bibinfo {year} {2009})},\ \Eprint
  {http://arxiv.org/abs/0811.1452} {arXiv:0811.1452 [gr-qc]} \BibitemShut
  {NoStop}%
%%CITATION = ARXIV:0811.1452;%%
\bibitem [{\citenamefont {Musco}\ and\ \citenamefont
  {Miller}(2013)}]{Musco:2012au}%
  \BibitemOpen
  \bibfield  {author} {\bibinfo {author} {\bibfnamefont {I.}~\bibnamefont
  {Musco}}\ and\ \bibinfo {author} {\bibfnamefont {J.~C.}\ \bibnamefont
  {Miller}},\ }\href {\doibase 10.1088/0264-9381/30/14/145009} {\bibfield
  {journal} {\bibinfo  {journal} {Class. Quant. Grav.}\ }\textbf {\bibinfo
  {volume} {30}},\ \bibinfo {pages} {145009} (\bibinfo {year} {2013})},\
  \Eprint {http://arxiv.org/abs/1201.2379} {arXiv:1201.2379 [gr-qc]}
  \BibitemShut {NoStop}%
%%CITATION = ARXIV:1201.2379;%%
\bibitem [{\citenamefont {K{\"u}hnel}\ \emph {et~al.}(2016)\citenamefont
  {K{\"u}hnel}, \citenamefont {Rampf},\ and\ \citenamefont
  {Sandstad}}]{Kuhnel:2015vtw}%
  \BibitemOpen
  \bibfield  {author} {\bibinfo {author} {\bibfnamefont {F.}~\bibnamefont
  {K{\"u}hnel}}, \bibinfo {author} {\bibfnamefont {C.}~\bibnamefont {Rampf}}, \
  and\ \bibinfo {author} {\bibfnamefont {M.}~\bibnamefont {Sandstad}},\ }\href
  {\doibase 10.1140/epjc/s10052-016-3945-8} {\bibfield  {journal} {\bibinfo
  {journal} {Eur. Phys. J.}\ }\textbf {\bibinfo {volume} {C76}},\ \bibinfo
  {pages} {93} (\bibinfo {year} {2016})},\ \Eprint
  {http://arxiv.org/abs/1512.00488} {arXiv:1512.00488 [astro-ph.CO]}
  \BibitemShut {NoStop}%
%%CITATION = ARXIV:1512.00488;%%
\bibitem [{\citenamefont {Gundlach}\ and\ \citenamefont
  {Martin-Garcia}(2007)}]{Gundlach:2007gc}%
  \BibitemOpen
  \bibfield  {author} {\bibinfo {author} {\bibfnamefont {C.}~\bibnamefont
  {Gundlach}}\ and\ \bibinfo {author} {\bibfnamefont {J.~M.}\ \bibnamefont
  {Martin-Garcia}},\ }\href {\doibase 10.12942/lrr-2007-5} {\bibfield
  {journal} {\bibinfo  {journal} {Living Rev. Rel.}\ }\textbf {\bibinfo
  {volume} {10}},\ \bibinfo {pages} {5} (\bibinfo {year} {2007})},\ \Eprint
  {http://arxiv.org/abs/0711.4620} {arXiv:0711.4620 [gr-qc]} \BibitemShut
  {NoStop}%
%%CITATION = ARXIV:0711.4620;%%
\bibitem [{\citenamefont {Shibata}\ and\ \citenamefont
  {Sasaki}(1999)}]{Shibata:1999zs}%
  \BibitemOpen
  \bibfield  {author} {\bibinfo {author} {\bibfnamefont {M.}~\bibnamefont
  {Shibata}}\ and\ \bibinfo {author} {\bibfnamefont {M.}~\bibnamefont
  {Sasaki}},\ }\href {\doibase 10.1103/PhysRevD.60.084002} {\bibfield
  {journal} {\bibinfo  {journal} {Phys. Rev.}\ }\textbf {\bibinfo {volume}
  {D60}},\ \bibinfo {pages} {084002} (\bibinfo {year} {1999})},\ \Eprint
  {http://arxiv.org/abs/gr-qc/9905064} {arXiv:gr-qc/9905064 [gr-qc]}
  \BibitemShut {NoStop}%
%%CITATION = GR-QC/9905064;%%
\bibitem [{\citenamefont {Harada}\ \emph {et~al.}(2013)\citenamefont {Harada},
  \citenamefont {Yoo},\ and\ \citenamefont {Kohri}}]{Harada:2013epa}%
  \BibitemOpen
  \bibfield  {author} {\bibinfo {author} {\bibfnamefont {T.}~\bibnamefont
  {Harada}}, \bibinfo {author} {\bibfnamefont {C.-M.}\ \bibnamefont {Yoo}}, \
  and\ \bibinfo {author} {\bibfnamefont {K.}~\bibnamefont {Kohri}},\ }\href
  {\doibase 10.1103/PhysRevD.88.084051, 10.1103/PhysRevD.89.029903} {\bibfield
  {journal} {\bibinfo  {journal} {Phys. Rev.}\ }\textbf {\bibinfo {volume}
  {D88}},\ \bibinfo {pages} {084051} (\bibinfo {year} {2013})},\ \bibinfo
  {note} {[Erratum: Phys. Rev.D89,no.2,029903(2014)]},\ \Eprint
  {http://arxiv.org/abs/1309.4201} {arXiv:1309.4201 [astro-ph.CO]} \BibitemShut
  {NoStop}%
%%CITATION = ARXIV:1309.4201;%%
\bibitem [{\citenamefont {Nakama}\ \emph
  {et~al.}(2014{\natexlab{a}})\citenamefont {Nakama}, \citenamefont {Harada},
  \citenamefont {Polnarev},\ and\ \citenamefont {Yokoyama}}]{Nakama:2013ica}%
  \BibitemOpen
  \bibfield  {author} {\bibinfo {author} {\bibfnamefont {T.}~\bibnamefont
  {Nakama}}, \bibinfo {author} {\bibfnamefont {T.}~\bibnamefont {Harada}},
  \bibinfo {author} {\bibfnamefont {A.~G.}\ \bibnamefont {Polnarev}}, \ and\
  \bibinfo {author} {\bibfnamefont {J.}~\bibnamefont {Yokoyama}},\ }\href
  {\doibase 10.1088/1475-7516/2014/01/037} {\bibfield  {journal} {\bibinfo
  {journal} {JCAP}\ }\textbf {\bibinfo {volume} {1401}},\ \bibinfo {pages}
  {037} (\bibinfo {year} {2014}{\natexlab{a}})},\ \Eprint
  {http://arxiv.org/abs/1310.3007} {arXiv:1310.3007 [gr-qc]} \BibitemShut
  {NoStop}%
%%CITATION = ARXIV:1310.3007;%%
\bibitem [{\citenamefont {Nakama}\ \emph {et~al.}(2017)\citenamefont {Nakama},
  \citenamefont {Silk},\ and\ \citenamefont {Kamionkowski}}]{Nakama:2016gzw}%
  \BibitemOpen
  \bibfield  {author} {\bibinfo {author} {\bibfnamefont {T.}~\bibnamefont
  {Nakama}}, \bibinfo {author} {\bibfnamefont {J.}~\bibnamefont {Silk}}, \ and\
  \bibinfo {author} {\bibfnamefont {M.}~\bibnamefont {Kamionkowski}},\ }\href
  {\doibase 10.1103/PhysRevD.95.043511} {\bibfield  {journal} {\bibinfo
  {journal} {Phys. Rev.}\ }\textbf {\bibinfo {volume} {D95}},\ \bibinfo {pages}
  {043511} (\bibinfo {year} {2017})},\ \Eprint
  {http://arxiv.org/abs/1612.06264} {arXiv:1612.06264 [astro-ph.CO]}
  \BibitemShut {NoStop}%
%%CITATION = ARXIV:1612.06264;%%
\bibitem [{\citenamefont {Young}\ \emph {et~al.}(2014)\citenamefont {Young},
  \citenamefont {Byrnes},\ and\ \citenamefont {Sasaki}}]{Young:2014ana}%
  \BibitemOpen
  \bibfield  {author} {\bibinfo {author} {\bibfnamefont {S.}~\bibnamefont
  {Young}}, \bibinfo {author} {\bibfnamefont {C.~T.}\ \bibnamefont {Byrnes}}, \
  and\ \bibinfo {author} {\bibfnamefont {M.}~\bibnamefont {Sasaki}},\ }\href
  {\doibase 10.1088/1475-7516/2014/07/045} {\bibfield  {journal} {\bibinfo
  {journal} {JCAP}\ }\textbf {\bibinfo {volume} {1407}},\ \bibinfo {pages}
  {045} (\bibinfo {year} {2014})},\ \Eprint {http://arxiv.org/abs/1405.7023}
  {arXiv:1405.7023 [gr-qc]} \BibitemShut {NoStop}%
%%CITATION = ARXIV:1405.7023;%%
\bibitem [{\citenamefont {Ade}\ \emph {et~al.}(2016)\citenamefont {Ade} \emph
  {et~al.}}]{Ade:2015xua}%
  \BibitemOpen
  \bibfield  {author} {\bibinfo {author} {\bibfnamefont {P.~A.~R.}\
  \bibnamefont {Ade}} \emph {et~al.} (\bibinfo {collaboration} {Planck}),\
  }\href {\doibase 10.1051/0004-6361/201525830} {\bibfield  {journal} {\bibinfo
   {journal} {Astron. Astrophys.}\ }\textbf {\bibinfo {volume} {594}},\
  \bibinfo {pages} {A13} (\bibinfo {year} {2016})},\ \Eprint
  {http://arxiv.org/abs/1502.01589} {arXiv:1502.01589 [astro-ph.CO]}
  \BibitemShut {NoStop}%
%%CITATION = ARXIV:1502.01589;%%
\bibitem [{\citenamefont {Carr}\ \emph {et~al.}(2010)\citenamefont {Carr},
  \citenamefont {Kohri}, \citenamefont {Sendouda},\ and\ \citenamefont
  {Yokoyama}}]{Carr:2009jm}%
  \BibitemOpen
  \bibfield  {author} {\bibinfo {author} {\bibfnamefont {B.~J.}\ \bibnamefont
  {Carr}}, \bibinfo {author} {\bibfnamefont {K.}~\bibnamefont {Kohri}},
  \bibinfo {author} {\bibfnamefont {Y.}~\bibnamefont {Sendouda}}, \ and\
  \bibinfo {author} {\bibfnamefont {J.}~\bibnamefont {Yokoyama}},\ }\href
  {\doibase 10.1103/PhysRevD.81.104019} {\bibfield  {journal} {\bibinfo
  {journal} {Phys. Rev.}\ }\textbf {\bibinfo {volume} {D81}},\ \bibinfo {pages}
  {104019} (\bibinfo {year} {2010})},\ \Eprint {http://arxiv.org/abs/0912.5297}
  {arXiv:0912.5297 [astro-ph.CO]} \BibitemShut {NoStop}%
%%CITATION = ARXIV:0912.5297;%%
\bibitem [{\citenamefont {Barnacka}\ \emph {et~al.}(2012)\citenamefont
  {Barnacka}, \citenamefont {Glicenstein},\ and\ \citenamefont
  {Moderski}}]{Barnacka:2012bm}%
  \BibitemOpen
  \bibfield  {author} {\bibinfo {author} {\bibfnamefont {A.}~\bibnamefont
  {Barnacka}}, \bibinfo {author} {\bibfnamefont {J.~F.}\ \bibnamefont
  {Glicenstein}}, \ and\ \bibinfo {author} {\bibfnamefont {R.}~\bibnamefont
  {Moderski}},\ }\href {\doibase 10.1103/PhysRevD.86.043001} {\bibfield
  {journal} {\bibinfo  {journal} {Phys. Rev.}\ }\textbf {\bibinfo {volume}
  {D86}},\ \bibinfo {pages} {043001} (\bibinfo {year} {2012})},\ \Eprint
  {http://arxiv.org/abs/1204.2056} {arXiv:1204.2056 [astro-ph.CO]} \BibitemShut
  {NoStop}%
%%CITATION = ARXIV:1204.2056;%%
\bibitem [{\citenamefont {Graham}\ \emph {et~al.}(2015)\citenamefont {Graham},
  \citenamefont {Rajendran},\ and\ \citenamefont {Varela}}]{Graham:2015apa}%
  \BibitemOpen
  \bibfield  {author} {\bibinfo {author} {\bibfnamefont {P.~W.}\ \bibnamefont
  {Graham}}, \bibinfo {author} {\bibfnamefont {S.}~\bibnamefont {Rajendran}}, \
  and\ \bibinfo {author} {\bibfnamefont {J.}~\bibnamefont {Varela}},\ }\href
  {\doibase 10.1103/PhysRevD.92.063007} {\bibfield  {journal} {\bibinfo
  {journal} {Phys. Rev.}\ }\textbf {\bibinfo {volume} {D92}},\ \bibinfo {pages}
  {063007} (\bibinfo {year} {2015})},\ \Eprint
  {http://arxiv.org/abs/1505.04444} {arXiv:1505.04444 [hep-ph]} \BibitemShut
  {NoStop}%
%%CITATION = ARXIV:1505.04444;%%
\bibitem [{\citenamefont {Griest}\ \emph {et~al.}(2013)\citenamefont {Griest},
  \citenamefont {Cieplak},\ and\ \citenamefont {Lehner}}]{Griest:2013esa}%
  \BibitemOpen
  \bibfield  {author} {\bibinfo {author} {\bibfnamefont {K.}~\bibnamefont
  {Griest}}, \bibinfo {author} {\bibfnamefont {A.~M.}\ \bibnamefont {Cieplak}},
  \ and\ \bibinfo {author} {\bibfnamefont {M.~J.}\ \bibnamefont {Lehner}},\
  }\href {\doibase 10.1103/PhysRevLett.111.181302} {\bibfield  {journal}
  {\bibinfo  {journal} {Phys. Rev. Lett.}\ }\textbf {\bibinfo {volume} {111}},\
  \bibinfo {pages} {181302} (\bibinfo {year} {2013})}\BibitemShut {NoStop}%
%%CITATION = PRLTA,111,181302;%%
\bibitem [{\citenamefont {Tisserand}\ \emph {et~al.}(2007)\citenamefont
  {Tisserand} \emph {et~al.}}]{Tisserand:2006zx}%
  \BibitemOpen
  \bibfield  {author} {\bibinfo {author} {\bibfnamefont {P.}~\bibnamefont
  {Tisserand}} \emph {et~al.} (\bibinfo {collaboration} {EROS-2}),\ }\href
  {\doibase 10.1051/0004-6361:20066017} {\bibfield  {journal} {\bibinfo
  {journal} {Astron. Astrophys.}\ }\textbf {\bibinfo {volume} {469}},\ \bibinfo
  {pages} {387} (\bibinfo {year} {2007})},\ \Eprint
  {http://arxiv.org/abs/astro-ph/0607207} {arXiv:astro-ph/0607207 [astro-ph]}
  \BibitemShut {NoStop}%
%%CITATION = ASTRO-PH/0607207;%%
\bibitem [{\citenamefont {Ricotti}\ \emph {et~al.}(2008)\citenamefont
  {Ricotti}, \citenamefont {Ostriker},\ and\ \citenamefont
  {Mack}}]{Ricotti:2007au}%
  \BibitemOpen
  \bibfield  {author} {\bibinfo {author} {\bibfnamefont {M.}~\bibnamefont
  {Ricotti}}, \bibinfo {author} {\bibfnamefont {J.~P.}\ \bibnamefont
  {Ostriker}}, \ and\ \bibinfo {author} {\bibfnamefont {K.~J.}\ \bibnamefont
  {Mack}},\ }\href {\doibase 10.1086/587831} {\bibfield  {journal} {\bibinfo
  {journal} {Astrophys. J.}\ }\textbf {\bibinfo {volume} {680}},\ \bibinfo
  {pages} {829} (\bibinfo {year} {2008})},\ \Eprint
  {http://arxiv.org/abs/0709.0524} {arXiv:0709.0524 [astro-ph]} \BibitemShut
  {NoStop}%
%%CITATION = ARXIV:0709.0524;%%
\bibitem [{\citenamefont {Niikura}\ \emph {et~al.}(2017)\citenamefont
  {Niikura}, \citenamefont {Takada}, \citenamefont {Yasuda}, \citenamefont
  {Lupton}, \citenamefont {Sumi}, \citenamefont {More}, \citenamefont {More},
  \citenamefont {Oguri},\ and\ \citenamefont {Chiba}}]{Niikura:2017zjd}%
  \BibitemOpen
  \bibfield  {author} {\bibinfo {author} {\bibfnamefont {H.}~\bibnamefont
  {Niikura}}, \bibinfo {author} {\bibfnamefont {M.}~\bibnamefont {Takada}},
  \bibinfo {author} {\bibfnamefont {N.}~\bibnamefont {Yasuda}}, \bibinfo
  {author} {\bibfnamefont {R.~H.}\ \bibnamefont {Lupton}}, \bibinfo {author}
  {\bibfnamefont {T.}~\bibnamefont {Sumi}}, \bibinfo {author} {\bibfnamefont
  {S.}~\bibnamefont {More}}, \bibinfo {author} {\bibfnamefont {A.}~\bibnamefont
  {More}}, \bibinfo {author} {\bibfnamefont {M.}~\bibnamefont {Oguri}}, \ and\
  \bibinfo {author} {\bibfnamefont {M.}~\bibnamefont {Chiba}},\ }\href@noop {}
  {\  (\bibinfo {year} {2017})},\ \Eprint {http://arxiv.org/abs/1701.02151}
  {arXiv:1701.02151 [astro-ph.CO]} \BibitemShut {NoStop}%
%%CITATION = ARXIV:1701.02151;%%
\bibitem [{\citenamefont {Chen}\ \emph {et~al.}(2016)\citenamefont {Chen},
  \citenamefont {Huang},\ and\ \citenamefont {Wang}}]{Chen:2016pud}%
  \BibitemOpen
  \bibfield  {author} {\bibinfo {author} {\bibfnamefont {L.}~\bibnamefont
  {Chen}}, \bibinfo {author} {\bibfnamefont {Q.-G.}\ \bibnamefont {Huang}}, \
  and\ \bibinfo {author} {\bibfnamefont {K.}~\bibnamefont {Wang}},\ }\href
  {\doibase 10.1088/1475-7516/2016/12/044} {\bibfield  {journal} {\bibinfo
  {journal} {JCAP}\ }\textbf {\bibinfo {volume} {1612}},\ \bibinfo {pages}
  {044} (\bibinfo {year} {2016})},\ \Eprint {http://arxiv.org/abs/1608.02174}
  {arXiv:1608.02174 [astro-ph.CO]} \BibitemShut {NoStop}%
%%CITATION = ARXIV:1608.02174;%%
\bibitem [{\citenamefont {Ali-Ha{\"i}moud}\ and\ \citenamefont
  {Kamionkowski}(2017)}]{Ali-Haimoud:2016mbv}%
  \BibitemOpen
  \bibfield  {author} {\bibinfo {author} {\bibfnamefont {Y.}~\bibnamefont
  {Ali-Ha{\"i}moud}}\ and\ \bibinfo {author} {\bibfnamefont {M.}~\bibnamefont
  {Kamionkowski}},\ }\href {\doibase 10.1103/PhysRevD.95.043534} {\bibfield
  {journal} {\bibinfo  {journal} {Phys. Rev.}\ }\textbf {\bibinfo {volume}
  {D95}},\ \bibinfo {pages} {043534} (\bibinfo {year} {2017})},\ \Eprint
  {http://arxiv.org/abs/1612.05644} {arXiv:1612.05644 [astro-ph.CO]}
  \BibitemShut {NoStop}%
%%CITATION = ARXIV:1612.05644;%%
\bibitem [{\citenamefont {Aloni}\ \emph {et~al.}(2017)\citenamefont {Aloni},
  \citenamefont {Blum},\ and\ \citenamefont {Flauger}}]{Blum:2016cjs}%
  \BibitemOpen
  \bibfield  {author} {\bibinfo {author} {\bibfnamefont {D.}~\bibnamefont
  {Aloni}}, \bibinfo {author} {\bibfnamefont {K.}~\bibnamefont {Blum}}, \ and\
  \bibinfo {author} {\bibfnamefont {R.}~\bibnamefont {Flauger}},\ }\href
  {\doibase 10.1088/1475-7516/2017/05/017} {\bibfield  {journal} {\bibinfo
  {journal} {JCAP}\ }\textbf {\bibinfo {volume} {1705}},\ \bibinfo {pages}
  {017} (\bibinfo {year} {2017})},\ \Eprint {http://arxiv.org/abs/1612.06811}
  {arXiv:1612.06811 [astro-ph.CO]} \BibitemShut {NoStop}%
%%CITATION = ARXIV:1612.06811;%%
\bibitem [{\citenamefont {Horowitz}(2016)}]{Horowitz:2016lib}%
  \BibitemOpen
  \bibfield  {author} {\bibinfo {author} {\bibfnamefont {B.}~\bibnamefont
  {Horowitz}},\ }\href@noop {} {\  (\bibinfo {year} {2016})},\ \Eprint
  {http://arxiv.org/abs/1612.07264} {arXiv:1612.07264 [astro-ph.CO]}
  \BibitemShut {NoStop}%
%%CITATION = ARXIV:1612.07264;%%
\bibitem [{\citenamefont {Moore}\ \emph {et~al.}(2015)\citenamefont {Moore},
  \citenamefont {Cole},\ and\ \citenamefont {Berry}}]{Moore:2014lga}%
  \BibitemOpen
  \bibfield  {author} {\bibinfo {author} {\bibfnamefont {C.~J.}\ \bibnamefont
  {Moore}}, \bibinfo {author} {\bibfnamefont {R.~H.}\ \bibnamefont {Cole}}, \
  and\ \bibinfo {author} {\bibfnamefont {C.~P.~L.}\ \bibnamefont {Berry}},\
  }\href {\doibase 10.1088/0264-9381/32/1/015014} {\bibfield  {journal}
  {\bibinfo  {journal} {Class. Quant. Grav.}\ }\textbf {\bibinfo {volume}
  {32}},\ \bibinfo {pages} {015014} (\bibinfo {year} {2015})},\ \Eprint
  {http://arxiv.org/abs/1408.0740} {arXiv:1408.0740 [gr-qc]} \BibitemShut
  {NoStop}%
%%CITATION = ARXIV:1408.0740;%%
\bibitem [{\citenamefont {Janssen}\ \emph {et~al.}(2015)\citenamefont {Janssen}
  \emph {et~al.}}]{Janssen:2014dka}%
  \BibitemOpen
  \bibfield  {author} {\bibinfo {author} {\bibfnamefont {G.}~\bibnamefont
  {Janssen}} \emph {et~al.},\ }\bibfield  {booktitle} {\emph {\bibinfo
  {booktitle} {{Proceedings, Advancing Astrophysics with the Square Kilometre
  Array (AASKA14): Giardini Naxos, Italy, June 9-13, 2014}}},\ }\href@noop {}
  {\bibfield  {journal} {\bibinfo  {journal} {PoS}\ }\textbf {\bibinfo {volume}
  {AASKA14}},\ \bibinfo {pages} {037} (\bibinfo {year} {2015})},\ \Eprint
  {http://arxiv.org/abs/1501.00127} {arXiv:1501.00127 [astro-ph.IM]}
  \BibitemShut {NoStop}%
%%CITATION = ARXIV:1501.00127;%%
\bibitem [{\citenamefont {Fixsen}\ \emph {et~al.}(1996)\citenamefont {Fixsen},
  \citenamefont {Cheng}, \citenamefont {Gales}, \citenamefont {Mather},
  \citenamefont {Shafer},\ and\ \citenamefont {Wright}}]{Fixsen:1996nj}%
  \BibitemOpen
  \bibfield  {author} {\bibinfo {author} {\bibfnamefont {D.~J.}\ \bibnamefont
  {Fixsen}}, \bibinfo {author} {\bibfnamefont {E.~S.}\ \bibnamefont {Cheng}},
  \bibinfo {author} {\bibfnamefont {J.~M.}\ \bibnamefont {Gales}}, \bibinfo
  {author} {\bibfnamefont {J.~C.}\ \bibnamefont {Mather}}, \bibinfo {author}
  {\bibfnamefont {R.~A.}\ \bibnamefont {Shafer}}, \ and\ \bibinfo {author}
  {\bibfnamefont {E.~L.}\ \bibnamefont {Wright}},\ }\href {\doibase
  10.1086/178173} {\bibfield  {journal} {\bibinfo  {journal} {Astrophys. J.}\
  }\textbf {\bibinfo {volume} {473}},\ \bibinfo {pages} {576} (\bibinfo {year}
  {1996})},\ \Eprint {http://arxiv.org/abs/astro-ph/9605054}
  {arXiv:astro-ph/9605054 [astro-ph]} \BibitemShut {NoStop}%
%%CITATION = ASTRO-PH/9605054;%%
\bibitem [{\citenamefont {Kogut}\ \emph {et~al.}(2011)\citenamefont {Kogut}
  \emph {et~al.}}]{Kogut:2011xw}%
  \BibitemOpen
  \bibfield  {author} {\bibinfo {author} {\bibfnamefont {A.}~\bibnamefont
  {Kogut}} \emph {et~al.},\ }\href {\doibase 10.1088/1475-7516/2011/07/025}
  {\bibfield  {journal} {\bibinfo  {journal} {JCAP}\ }\textbf {\bibinfo
  {volume} {1107}},\ \bibinfo {pages} {025} (\bibinfo {year} {2011})},\ \Eprint
  {http://arxiv.org/abs/1105.2044} {arXiv:1105.2044 [astro-ph.CO]} \BibitemShut
  {NoStop}%
%%CITATION = ARXIV:1105.2044;%%
\bibitem [{\citenamefont {Andre}\ \emph {et~al.}(2013)\citenamefont {Andre}
  \emph {et~al.}}]{Andre:2013afa}%
  \BibitemOpen
  \bibfield  {author} {\bibinfo {author} {\bibfnamefont {P.}~\bibnamefont
  {Andre}} \emph {et~al.} (\bibinfo {collaboration} {PRISM}),\ }\href@noop {}
  {\  (\bibinfo {year} {2013})},\ \Eprint {http://arxiv.org/abs/1306.2259}
  {arXiv:1306.2259 [astro-ph.CO]} \BibitemShut {NoStop}%
%%CITATION = ARXIV:1306.2259;%%
\bibitem [{\citenamefont {Jeong}\ \emph {et~al.}(2014)\citenamefont {Jeong},
  \citenamefont {Pradler}, \citenamefont {Chluba},\ and\ \citenamefont
  {Kamionkowski}}]{Jeong:2014gna}%
  \BibitemOpen
  \bibfield  {author} {\bibinfo {author} {\bibfnamefont {D.}~\bibnamefont
  {Jeong}}, \bibinfo {author} {\bibfnamefont {J.}~\bibnamefont {Pradler}},
  \bibinfo {author} {\bibfnamefont {J.}~\bibnamefont {Chluba}}, \ and\ \bibinfo
  {author} {\bibfnamefont {M.}~\bibnamefont {Kamionkowski}},\ }\href {\doibase
  10.1103/PhysRevLett.113.061301} {\bibfield  {journal} {\bibinfo  {journal}
  {Phys. Rev. Lett.}\ }\textbf {\bibinfo {volume} {113}},\ \bibinfo {pages}
  {061301} (\bibinfo {year} {2014})},\ \Eprint {http://arxiv.org/abs/1403.3697}
  {arXiv:1403.3697 [astro-ph.CO]} \BibitemShut {NoStop}%
%%CITATION = ARXIV:1403.3697;%%
\bibitem [{\citenamefont {Nakama}\ \emph
  {et~al.}(2014{\natexlab{b}})\citenamefont {Nakama}, \citenamefont {Suyama},\
  and\ \citenamefont {Yokoyama}}]{Nakama:2014vla}%
  \BibitemOpen
  \bibfield  {author} {\bibinfo {author} {\bibfnamefont {T.}~\bibnamefont
  {Nakama}}, \bibinfo {author} {\bibfnamefont {T.}~\bibnamefont {Suyama}}, \
  and\ \bibinfo {author} {\bibfnamefont {J.}~\bibnamefont {Yokoyama}},\ }\href
  {\doibase 10.1103/PhysRevLett.113.061302} {\bibfield  {journal} {\bibinfo
  {journal} {Phys. Rev. Lett.}\ }\textbf {\bibinfo {volume} {113}},\ \bibinfo
  {pages} {061302} (\bibinfo {year} {2014}{\natexlab{b}})},\ \Eprint
  {http://arxiv.org/abs/1403.5407} {arXiv:1403.5407 [astro-ph.CO]} \BibitemShut
  {NoStop}%
%%CITATION = ARXIV:1403.5407;%%
\bibitem [{\citenamefont {Inomata}\ \emph {et~al.}(2016)\citenamefont
  {Inomata}, \citenamefont {Kawasaki},\ and\ \citenamefont
  {Tada}}]{Inomata:2016uip}%
  \BibitemOpen
  \bibfield  {author} {\bibinfo {author} {\bibfnamefont {K.}~\bibnamefont
  {Inomata}}, \bibinfo {author} {\bibfnamefont {M.}~\bibnamefont {Kawasaki}}, \
  and\ \bibinfo {author} {\bibfnamefont {Y.}~\bibnamefont {Tada}},\ }\href
  {\doibase 10.1103/PhysRevD.94.043527} {\bibfield  {journal} {\bibinfo
  {journal} {Phys. Rev.}\ }\textbf {\bibinfo {volume} {D94}},\ \bibinfo {pages}
  {043527} (\bibinfo {year} {2016})},\ \Eprint
  {http://arxiv.org/abs/1605.04646} {arXiv:1605.04646 [astro-ph.CO]}
  \BibitemShut {NoStop}%
%%CITATION = ARXIV:1605.04646;%%
\bibitem [{\citenamefont {Izawa}\ \emph {et~al.}(1997)\citenamefont {Izawa},
  \citenamefont {Kawasaki},\ and\ \citenamefont {Yanagida}}]{Izawa:1997df}%
  \BibitemOpen
  \bibfield  {author} {\bibinfo {author} {\bibfnamefont {K.~I.}\ \bibnamefont
  {Izawa}}, \bibinfo {author} {\bibfnamefont {M.}~\bibnamefont {Kawasaki}}, \
  and\ \bibinfo {author} {\bibfnamefont {T.}~\bibnamefont {Yanagida}},\ }\href
  {\doibase 10.1016/S0370-2693(97)01040-X} {\bibfield  {journal} {\bibinfo
  {journal} {Phys. Lett.}\ }\textbf {\bibinfo {volume} {B411}},\ \bibinfo
  {pages} {249} (\bibinfo {year} {1997})},\ \Eprint
  {http://arxiv.org/abs/hep-ph/9707201} {arXiv:hep-ph/9707201 [hep-ph]}
  \BibitemShut {NoStop}%
%%CITATION = HEP-PH/9707201;%%
\bibitem [{\citenamefont {Weinberg}(2008)}]{Weinberg:2008zzc}%
  \BibitemOpen
  \bibfield  {author} {\bibinfo {author} {\bibfnamefont {S.}~\bibnamefont
  {Weinberg}},\ }\href {http://www.oup.com/uk/catalogue/?ci=9780198526827}
  {\emph {\bibinfo {title} {{Cosmology}}}}\ (\bibinfo {year}
  {2008})\BibitemShut {NoStop}%
%%CITATION = INSPIRE-794379;%%
\bibitem [{\citenamefont {Ema}\ \emph {et~al.}(2016)\citenamefont {Ema},
  \citenamefont {Mukaida},\ and\ \citenamefont {Nakayama}}]{Ema:2016kpf}%
  \BibitemOpen
  \bibfield  {author} {\bibinfo {author} {\bibfnamefont {Y.}~\bibnamefont
  {Ema}}, \bibinfo {author} {\bibfnamefont {K.}~\bibnamefont {Mukaida}}, \ and\
  \bibinfo {author} {\bibfnamefont {K.}~\bibnamefont {Nakayama}},\ }\href
  {\doibase 10.1088/1475-7516/2016/10/043} {\bibfield  {journal} {\bibinfo
  {journal} {JCAP}\ }\textbf {\bibinfo {volume} {1610}},\ \bibinfo {pages}
  {043} (\bibinfo {year} {2016})},\ \Eprint {http://arxiv.org/abs/1602.00483}
  {arXiv:1602.00483 [hep-ph]} \BibitemShut {NoStop}%
%%CITATION = ARXIV:1602.00483;%%
\bibitem [{\citenamefont {Saito}\ \emph {et~al.}(2008)\citenamefont {Saito},
  \citenamefont {Yokoyama},\ and\ \citenamefont {Nagata}}]{Saito:2008em}%
  \BibitemOpen
  \bibfield  {author} {\bibinfo {author} {\bibfnamefont {R.}~\bibnamefont
  {Saito}}, \bibinfo {author} {\bibfnamefont {J.}~\bibnamefont {Yokoyama}}, \
  and\ \bibinfo {author} {\bibfnamefont {R.}~\bibnamefont {Nagata}},\ }\href
  {\doibase 10.1088/1475-7516/2008/06/024} {\bibfield  {journal} {\bibinfo
  {journal} {JCAP}\ }\textbf {\bibinfo {volume} {0806}},\ \bibinfo {pages}
  {024} (\bibinfo {year} {2008})},\ \Eprint {http://arxiv.org/abs/0804.3470}
  {arXiv:0804.3470 [astro-ph]} \BibitemShut {NoStop}%
%%CITATION = ARXIV:0804.3470;%%
\bibitem [{\citenamefont {Kawasaki}\ \emph {et~al.}(2013)\citenamefont
  {Kawasaki}, \citenamefont {Kitajima},\ and\ \citenamefont
  {Yanagida}}]{Kawasaki:2012wr}%
  \BibitemOpen
  \bibfield  {author} {\bibinfo {author} {\bibfnamefont {M.}~\bibnamefont
  {Kawasaki}}, \bibinfo {author} {\bibfnamefont {N.}~\bibnamefont {Kitajima}},
  \ and\ \bibinfo {author} {\bibfnamefont {T.~T.}\ \bibnamefont {Yanagida}},\
  }\href {\doibase 10.1103/PhysRevD.87.063519} {\bibfield  {journal} {\bibinfo
  {journal} {Phys. Rev.}\ }\textbf {\bibinfo {volume} {D87}},\ \bibinfo {pages}
  {063519} (\bibinfo {year} {2013})},\ \Eprint {http://arxiv.org/abs/1207.2550}
  {arXiv:1207.2550 [hep-ph]} \BibitemShut {NoStop}%
%%CITATION = ARXIV:1207.2550;%%
\bibitem [{\citenamefont {Kohri}\ \emph {et~al.}(2013)\citenamefont {Kohri},
  \citenamefont {Lin},\ and\ \citenamefont {Matsuda}}]{Kohri:2012yw}%
  \BibitemOpen
  \bibfield  {author} {\bibinfo {author} {\bibfnamefont {K.}~\bibnamefont
  {Kohri}}, \bibinfo {author} {\bibfnamefont {C.-M.}\ \bibnamefont {Lin}}, \
  and\ \bibinfo {author} {\bibfnamefont {T.}~\bibnamefont {Matsuda}},\ }\href
  {\doibase 10.1103/PhysRevD.87.103527} {\bibfield  {journal} {\bibinfo
  {journal} {Phys. Rev.}\ }\textbf {\bibinfo {volume} {D87}},\ \bibinfo {pages}
  {103527} (\bibinfo {year} {2013})},\ \Eprint {http://arxiv.org/abs/1211.2371}
  {arXiv:1211.2371 [hep-ph]} \BibitemShut {NoStop}%
%%CITATION = ARXIV:1211.2371;%%
\bibitem [{\citenamefont {Garcia-Bellido}\ \emph {et~al.}(2016)\citenamefont
  {Garcia-Bellido}, \citenamefont {Peloso},\ and\ \citenamefont
  {Unal}}]{Garcia-Bellido:2016dkw}%
  \BibitemOpen
  \bibfield  {author} {\bibinfo {author} {\bibfnamefont {J.}~\bibnamefont
  {Garcia-Bellido}}, \bibinfo {author} {\bibfnamefont {M.}~\bibnamefont
  {Peloso}}, \ and\ \bibinfo {author} {\bibfnamefont {C.}~\bibnamefont
  {Unal}},\ }\href {\doibase 10.1088/1475-7516/2016/12/031} {\bibfield
  {journal} {\bibinfo  {journal} {JCAP}\ }\textbf {\bibinfo {volume} {1612}},\
  \bibinfo {pages} {031} (\bibinfo {year} {2016})},\ \Eprint
  {http://arxiv.org/abs/1610.03763} {arXiv:1610.03763 [astro-ph.CO]}
  \BibitemShut {NoStop}%
%%CITATION = ARXIV:1610.03763;%%
\bibitem [{\citenamefont {Cheng}\ \emph {et~al.}(2017)\citenamefont {Cheng},
  \citenamefont {Lee},\ and\ \citenamefont {Ng}}]{Cheng:2016qzb}%
  \BibitemOpen
  \bibfield  {author} {\bibinfo {author} {\bibfnamefont {S.-L.}\ \bibnamefont
  {Cheng}}, \bibinfo {author} {\bibfnamefont {W.}~\bibnamefont {Lee}}, \ and\
  \bibinfo {author} {\bibfnamefont {K.-W.}\ \bibnamefont {Ng}},\ }\href
  {\doibase 10.1007/JHEP02(2017)008} {\bibfield  {journal} {\bibinfo  {journal}
  {JHEP}\ }\textbf {\bibinfo {volume} {02}},\ \bibinfo {pages} {008} (\bibinfo
  {year} {2017})},\ \Eprint {http://arxiv.org/abs/1606.00206} {arXiv:1606.00206
  [astro-ph.CO]} \BibitemShut {NoStop}%
%%CITATION = ARXIV:1606.00206;%%
\bibitem [{\citenamefont {Kawasaki}\ \emph {et~al.}(2000)\citenamefont
  {Kawasaki}, \citenamefont {Yamaguchi},\ and\ \citenamefont
  {Yanagida}}]{Kawasaki:2000yn}%
  \BibitemOpen
  \bibfield  {author} {\bibinfo {author} {\bibfnamefont {M.}~\bibnamefont
  {Kawasaki}}, \bibinfo {author} {\bibfnamefont {M.}~\bibnamefont {Yamaguchi}},
  \ and\ \bibinfo {author} {\bibfnamefont {T.}~\bibnamefont {Yanagida}},\
  }\href {\doibase 10.1103/PhysRevLett.85.3572} {\bibfield  {journal} {\bibinfo
   {journal} {Phys. Rev. Lett.}\ }\textbf {\bibinfo {volume} {85}},\ \bibinfo
  {pages} {3572} (\bibinfo {year} {2000})},\ \Eprint
  {http://arxiv.org/abs/hep-ph/0004243} {arXiv:hep-ph/0004243 [hep-ph]}
  \BibitemShut {NoStop}%
%%CITATION = HEP-PH/0004243;%%
\end{thebibliography}%
\end{document}